\documentclass[aps,prc,preprint,showpacs, reprint, showkeys, amsmath,amssymb,groupedaddress]{revtex4-1}

\usepackage{hyperref}
\usepackage{graphicx} 
\usepackage{dcolumn}     
\usepackage{bm}   
\usepackage{color}  
 
\begin{document}  


\title{Polarized e-p Elastic Scattering in the Collider Frame} 


\author{C. Sofiatti}
\affiliation{Department of Physics,\\ 
University of Massachusetts Boston,  
100 Morrissey Blvd., Boston, MA 02125}  
\email {c.sofiatti@gmail.com}

\author{T.~W.~Donnelly}
\affiliation{Center for Theoretical Physics, Laboratory for Nuclear
Science and Department of Physics, Massachusetts Institute of
Technology, Cambridge, MA 02139}


\date{\today}


\begin{abstract}
Double polarization elastic $\vec{e}$-$\vec{p}$ cross sections and
asymmetries are considered in collider kinematics. Covariant
expressions are derived for the general situation involving crossed
beams; these are checked against the well-known results obtained
when the proton is at rest. Results are given using modern models
for the proton electromagnetic form factors for kinematics of
interest in e-p colliders such as the EIC facility which is in its
planning stage. In context, parity-violating elastic $\vec{e}$-$p$
scattering is compared and contrasted with these double-polarization
(parity-conserving) results.
\end{abstract}




\maketitle


\section{Introduction\label{sec:intro}}

The main focus of this paper is the development of the formalism for
polarized $\vec{e}$-$\vec{p}$ elastic scattering in the collider
frame, stimulated by interest in e-p colliders such as the EIC
\cite{EIC} facility which is in its planning stage. At least two
things motivate such a study. (1) One has the possibility of
measuring the electromagnetic form factors of the proton in unusual
kinematics, that is, with high-energy colliding beams of polarized
electrons and protons --- to be contrasted to the usual situation
with polarized electrons scattering from protons at rest with either
the target proton polarized or when the recoiling final-state
proton's polarization is measured. Specifically, we shall see that
the role played by $2\gamma$ corrections to the dominantly
one-photon-exchange diagram (the only one we consider in detail in
this work) is likely quite different for collider kinematics. (2)
One may use the reasonably well-known double-polarization asymmetry
to determine the product of the electron and proton polarizations,
$p_e p_p$, when the focus is placed on other reactions ({\it e.g.,}
DIS).

The basic kinematical formalism is presented in Sect.\ \ref{sec:kine}
for both collinear and crossed beams. This is followed in Sect.\ \ref{sec:tensors} by developments of the leptonic (electron) and
hadronic (proton) tensors needed in discussing parity-conserving
double-polarized $\vec{e}$-$\vec{p}$ elastic scattering. Here a
general frame is considered and all quantities are kept completely
covariant so that any situation can be explored, including the
special case where the proton is at rest (the ``rest frame'') and
where the formalism is well known. In Sect.\ \ref{sec:contract} the
leptonic and hadronic tensors are contracted to yield the invariant
matrix element required in constructing the polarized cross sections
and asymmetries, while in Sect.\ \ref{sec:cross-asym} expressions for
the latter are presented.

Our approach in this study has been to develop the formalism in
detail and thereby to bring out clearly the particular roles played
by the different proton form factors. We shall see that the fact
that the proton's Pauli form factor is nonzero alters the character
of the asymmetry from the simpler answer obtained when colliding
point Dirac particles. We retain both the electron and proton masses
({\it i.e.,} we do not invoke the extreme relativistic limit) so
that any choice of kinematics can be explored. In the case of the
electron this is not needed except when scattering at very small
angles; however, for the proton it is critical to retain the mass
terms if one wishes to be able to go to the rest frame. Since the
formalism is only a little more involved when keeping all mass terms
than to drop them as is often done, we retain them throughout this
study.

We shall see that the asymmetries are rather small --- the reasons
for this will be explained later --- and accordingly in Sect.\ \ref{sec:PV} we also briefly consider parity-violating elastic
$\vec{e}$-$p$ scattering in collider kinematics. We shall see that
this single-polarization asymmetry is only about one order of
magnitude smaller than the double-polarization parity-conserving
asymmetries.

In Sect.\ \ref{sec:numerics} results are presented for two choices of
kinematics that may be relevant for a future EIC facility. The
asymmetries (PC and PV) are all given, as is the figure-of-merit and
thereby the anticipated fractional uncertainty expected given
specific collider luminosities and polarizations. The computer code
Brasil2011 has been developed to handle any kinematical situation
and can be obtained \footnote{ The C++ computer code Brasil2011 that yields all of the kinematic variables, cross sections, asymmetries and figures-of-merit may be obtained by contacting c.sofiatti@gmail.com.} by anyone interested in exploring
other conditions that may be relevant when planning for a future e-p
collider. Finally, our conclusions are presented in Sect.
\ref{sec:concl}.


\section{Basic Collider-Frame Kinematics\label{sec:kine}}


\subsection{Collinear beams\label{subsec:collinear}}


\begin{figure}[hbt]
  \includegraphics[width=0.47\textwidth,viewport=25 154 831 493]{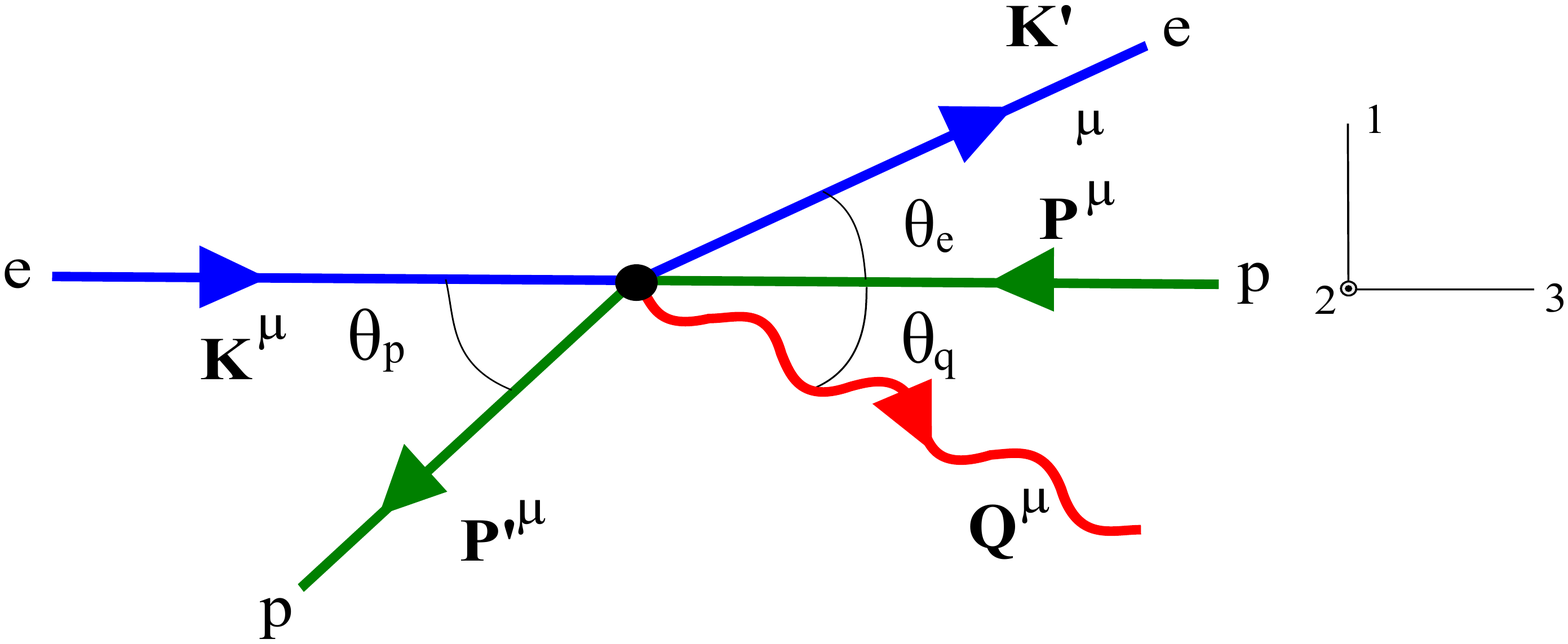}
  \caption{(color online) Electron-proton elastic scattering in collider kinematics.}
  \label{fig:fig1}
\end{figure}


The coordinate frame used in this work is shown in Fig.
\ref{fig:fig1}; we start with collinear kinematics and then in the
following subsection generalize to the situation where the electron
and proton beams are crossed. Here an electron with 4-momentum
$K^{\mu }$ is incident from the left and a proton with $P^{\mu }$
enters from the right. The final state has an electron with
$K^{\prime \mu }$ scattered at an angle $\theta
_{e}$ and a proton with $P^{\prime \mu }$ scattered at an angle $%
\theta _{p}$, as shown. The 4-momentum transfer
\begin{equation}
Q^{\mu }=K^{\mu }-K^{\prime \mu }=P^{\prime \mu }-P^{\mu }
\label{eq5}
\end{equation}%
makes an angle $\theta _{q}$ with respect to the beam axis. The
4-momenta in the problem are thus the following: $K^{\mu
}=(\epsilon ,\mathbf{k})$, $K^{\prime \mu }=(\epsilon ^{\prime },\mathbf{k}%
^{\prime })$, $P^{\mu }=(E,\mathbf{p})$, $P^{\prime \mu }=(E^{\prime },%
\mathbf{p}^{\prime })$ and $Q^{\mu }=(\omega ,\mathbf{q})$ with $\mathbf{k}=k%
\mathbf{u}_{3}$ and $\mathbf{p}=-p\mathbf{u}_{3}$. Since the
electron and proton are both on-shell, one can write their energies
in terms of their
3-momenta: $\epsilon =\sqrt{k^{2}+m_{e}^{2}}$, $\epsilon ^{\prime }=\sqrt{%
k^{\prime 2}+m_{e}^{2}}$, $E=\sqrt{p^{2}+m_{p}^{2}}$ and $E^{\prime }=\sqrt{%
p^{\prime 2}+m_{p}^{2}}$.

We assume that the variables used to specify the kinematics are
$(k,p,\theta _{e})$ and then through the equations above the
energies of the incident particles are also given. It proves useful
to define the total 4-momentum
\begin{eqnarray}
P_{tot}^{\mu } &\equiv &K^{\mu }+P^{\mu }\equiv
(E_{tot},\mathbf{p}_{tot})
\label{eq23} \\
&=&K^{\prime \mu }+P^{\prime \mu },  \label{eq24}
\end{eqnarray}%
where Eq.\ (\ref{eq24}) follows by energy-momentum conservation. This
implies that
\begin{eqnarray}
E_{tot} &=&\epsilon +E=\epsilon ^{\prime }+E^{\prime }  \label{eq25} \\
\mathbf{p}_{tot} &=&(k-p)\mathbf{u}_{3}=p_{tot}\mathbf{u}_{3}=\mathbf{k}%
^{\prime }+\mathbf{p}^{\prime }.  \label{eq26}
\end{eqnarray}%
The square of the total 4-momentum is the invariant
\begin{equation}
s=P_{tot}^{2}=E_{tot}^{2}-p_{tot}^{2}=m_{e}^{2}+m_{p}^{2}+2\xi ,
\label{eq29}
\end{equation}%
where
\begin{equation}
\xi \equiv \epsilon E+kp  \label{eq30}
\end{equation}%
is also given by the initial momenta.

Using 3-momentum conservation and given $p_{tot}$ and $\theta _{e}$,
the scattered proton's 3-momentum can be written in terms of the
scattered electron's 3-momentum:
\begin{eqnarray}
p^{\prime } &=&\sqrt{k^{\prime 2}+p_{tot}^{2}-2p_{tot}k^{\prime
}\cos \theta
_{e}}  \label{eq36} \\
\sin \theta _{p} &=&\frac{k^{\prime }}{p^{\prime }}\sin \theta _{e}
\label{eq37} \\
\cos \theta _{p} &=&\frac{1}{p^{\prime }}\left( k^{\prime }\cos
\theta _{e}-p_{tot}\right) .  \label{eq38}
\end{eqnarray}%
Additionally, from energy conservation one has that
\begin{equation}
E_{tot}\epsilon ^{\prime }-p_{tot}k^{\prime }\cos \theta
_{e}=m_{e}^{2}+\xi \label{eq43}
\end{equation}%
which constitutes an equation for $k^{\prime }$. Solving, one has
\begin{equation}
k^{\prime }=\frac{1}{a}\left[ b+E_{tot}\sqrt{\xi
^{2}-m_{e}^{2}(m_{p}^{2}+p_{tot}^{2}\sin ^{2}\theta _{e})}\right]
\label{eq44}
\end{equation}%
with
\begin{eqnarray}
a &\equiv &E_{tot}^{2}-p_{tot}^{2}\cos ^{2}\theta
_{e}=s+p_{tot}^{2}\sin
^{2}\theta _{e}  \label{eq45} \\
&=&m_{p}^{2}+m_{e}^{2}+2\xi +p_{tot}^{2}\sin ^{2}\theta _{e} \nonumber\\
&&\qquad\qquad\geq (m_{p}+m_{e})^{2}>0 \label{eq45a} \\
b &\equiv &(m_{e}^{2}+\xi )p_{tot}\cos \theta _{e},  \label{eq46}
\end{eqnarray}%
and knowing $k^{\prime }$ one can use the equations given above to
determine $\epsilon ^{\prime },p^{\prime },E^{\prime }$ and $\theta 
_{p}$. Since the argument of the square root in Eq.\ (\ref{eq44})
must be non-negative one has
\begin{equation}
\xi \geq m_{e}\sqrt{m_{p}^{2}+p_{tot}^{2}\sin ^{2}\theta _{e}}.
\label{eq47}
\end{equation}%
The 4-momentum transfer is also now specified:
\begin{eqnarray}
\omega &=&\epsilon -\epsilon ^{\prime }  \label{eq48} \\
q &=&\sqrt{k^{2}+k^{\prime 2}-2kk^{\prime }\cos \theta _{e}}  \label{eq49} \\
\sin \theta _{q} &=&\frac{k^{\prime }}{q}\sin \theta _{e}  \label{eq50} \\
\cos \theta _{q} &=&\frac{1}{q}\left( k-k^{\prime }\cos \theta
_{e}\right) . \label{eq51}
\end{eqnarray}%
From Eqs. (\ref{eq48}) and (\ref{eq49}), together with the
energy-momentum relationships above one has
\begin{eqnarray}
Q^{2} &=&t=-2(\epsilon \epsilon ^{\prime }-kk^{\prime }\cos \theta
_{e}-m_{e}^{2})  \label{eq56} \\
&=&-\left[ 4kk^{\prime }\sin ^{2}\theta
_{e}/2+\frac{2m_{e}^{2}(k-k^{\prime })^{2}}{\epsilon \epsilon
^{\prime }+kk^{\prime }+m_{e}^{2}}\right] \leq 0, \label{eq57}
\end{eqnarray}%
that is, the 4-momentum transfer is spacelike\ \footnote{%
NB: in the conventions employed in this and other work upon which
these studies are based the 4-vector conventions outlined in the Appendix are adopted and consequently $Q^{2}$ is negative when
spacelike.}.

To conclude this brief discussion of the basic collinear kinematics,
it is instructive to express the kinematic variables above in
the proton rest frame. For any 4-vector in the collider frame the
corresponding
quantities in the proton rest frame may be found by boosting in the $\mathbf{%
u}_{3}$ direction by $\beta _{p}\equiv p/E$ with $\gamma
_{p}=E/m_{p}=[1-\beta _{p}^{2}]^{-1/2}$. In particular, the proton
in its rest frame of course has $P^{\mu }=(m_{p},0,0,0)$ while the
incident electron has 3-momentum and energy given by
\begin{eqnarray}
k_{rest} &=&\gamma _{p}(k+\beta _{p}\epsilon )  \label{eqrest5} \\
\epsilon _{rest} &=&\gamma _{p}(\epsilon +\beta _{p}k).
\label{eqrest6}
\end{eqnarray}%
For the scattered electron one has
\begin{eqnarray}
\epsilon _{rest}^{\prime } &=&\gamma _{p}(\epsilon ^{\prime }+\beta
_{p}k^{\prime }\cos \theta _{e})  \label{eqrest7} \\
k_{rest}^{\prime } &=&\sqrt{\left( \epsilon _{rest}^{\prime }\right)
^{2}-m_{e}^{2}}  \label{eqrest8} \\
\sin \theta _{e}^{rest} &=&\frac{k^{\prime }\sin \theta _{e}}{%
k_{rest}^{\prime }}  \label{eqrest9} \\
\cos \theta _{e}^{rest} &=&\frac{\gamma _{p}(k^{\prime }\cos \theta
_{e}+\beta _{p}\epsilon ^{\prime })}{k_{rest}^{\prime }}.
\label{eqrest10}
\end{eqnarray}

To get some feeling for the extreme nature of the kinematics
typically of interest (see Sect.\ \ref{sec:numerics}), we consider two
choices for the kinematics, I. $k=10$ GeV/c with $p=250$ GeV/c and
II. $k=2$ GeV/c with $p=50$ GeV/c (see also Table I in Sect.\ \ref{sec:numerics}). For the former high-energy case
we have $\beta _{p}\approx 1$, $\gamma _{p}\approx 250$, so $%
k_{rest}\approx 500k=5$ TeV/c and $\sin \theta _{e}^{rest}\approx
\sin \theta _{e}/500$. This means that for this choice of kinematics
and, for instance, for $1^{\circ}$ ($5^{\circ}$) scattering in the collider
frame the equivalent rest frame (for example, for fixed target
measurements) has an incident electron beam of $5.3$ TeV/c
scattering at $0.002^{\circ}$ ($0.01^{\circ}$). For the lower-energy case
II. we have again for $1^{\circ}$ ($5^{\circ}$) scattering in the collider
frame that the equivalent rest frame has an incident electron beam
of $213$ GeV/c with scattering angle $0.009^{\circ}$ ($0.047^{\circ}$).

Any of the other kinematic variables above may be related to their
rest-frame equivalents in a similar manner.


\subsection{Crossed beams\label{subsec:crossed}}


\begin{figure}[hbt]
  \includegraphics[width=0.47\textwidth,viewport=21 121 826 487]{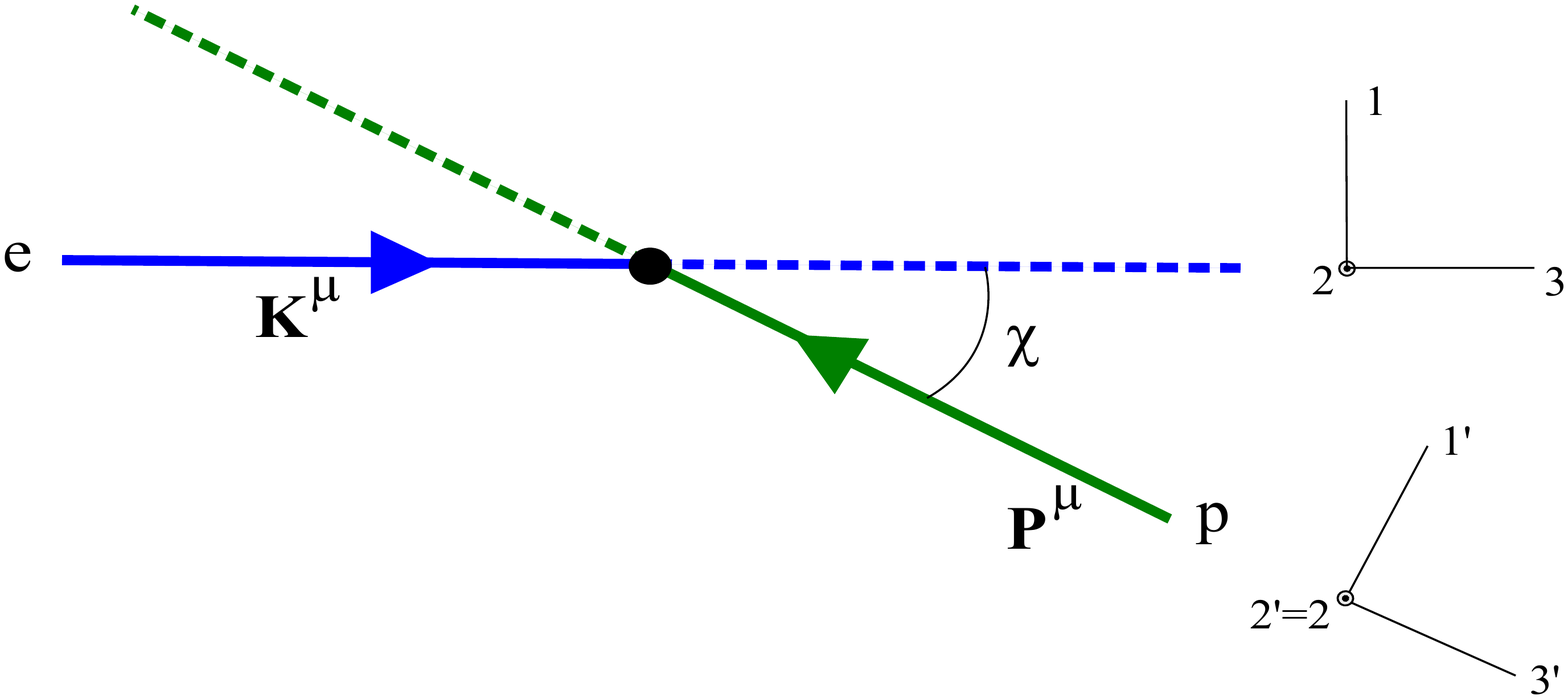}
  \caption{(color online)  Electron-proton elastic scattering in crossed-beams
collider kinematics.}
  \label{fig:fig2}
\end{figure}


In this subsection we provide the extensions that are necessary when
the electron and proton beams are not collinear, but are crossed.
The kinematics for this are shown in Fig.\ \ref{fig:fig2}. The
electron is assumed to be incident along the $\mathbf{u}_{3}$-axis
as before; however, now the proton beam is assumed to have its
momentum $\mathbf{p}$ directed along the $-\mathbf{u}_{3^{\prime }}$%
-axis where the ($1^{\prime },2^{\prime },3^{\prime }$) system is rotated from the (%
$1,2,3$) through an angle $\chi $ as shown. Clearly%
\begin{eqnarray}
\mathbf{u}_{1} &=&\cos \chi \mathbf{u}_{1^{\prime }}-\sin \chi \mathbf{u}%
_{3^{\prime }}  \label{eqz3} \\
\mathbf{u}_{3} &=&\sin \chi \mathbf{u}_{1^{\prime }}+\cos \chi \mathbf{u}%
_{3^{\prime }}.  \label{eqz4}
\end{eqnarray}%
Let us repeat the kinematics developments of the previous
subsection, now
working in the ($1^{\prime },2^{\prime },3^{\prime }$) system where we have $%
\mathbf{k}=k\left( \sin \chi \mathbf{u}_{1^{\prime }}+\cos \chi \mathbf{u}%
_{3^{\prime }}\right) $ and $\mathbf{p}=-p\mathbf{u}_{3^{\prime }}$.
$E_{tot}
$ in Eq.\ (\ref{eq25}) is as before; however, now we have%
\begin{equation}
\mathbf{p}_{tot}=k\sin \chi \mathbf{u}_{1^{\prime }}+\left( k\cos
\chi -p\right) \mathbf{u}_{3^{\prime }},  \label{eqz9}
\end{equation}%
yielding%
\begin{equation}
p_{tot}^{2}=k^{2}+p^{2}-2kp\cos \chi   \label{eqz10}
\end{equation}%
and therefore%
\begin{equation}
s=P_{tot}^{2}=E_{tot}^{2}-p_{tot}^{2}=m_{e}^{2}+m_{p}^{2}+2\widehat{\xi
}, \label{eqz11}
\end{equation}%
where%
\begin{equation}
\widehat{\xi }=\epsilon E+kp\cos \chi ;  \label{eqz12}
\end{equation}%
\textit{cf.} Eq.\ (\ref{eq30}). The extension of Eq.\ (\ref{eq36}) is%
\begin{equation}
p^{\prime }=\sqrt{k^{\prime 2}+p_{tot}^{2}-2k^{\prime }\left( k\cos
\theta _{e}-p\cos \left( \theta _{e}+\chi \right) \right) }
\label{eqz19}
\end{equation}%
and of Eqs. (\ref{eq37},\ref{eq38}) are%
\begin{eqnarray}
\sin \left( \theta _{p}+\chi \right)  &=&\frac{1}{p^{\prime }}\left[
k^{\prime }\sin \left( \theta _{e}+\chi \right) -k\sin \chi \right]
\label{eqz20} \\
\cos \left( \theta _{p}+\chi \right)  &=&\frac{1}{p^{\prime }}\left[
k^{\prime }\cos \left( \theta _{e}+\chi \right) -\left( k\cos \chi -p\right) %
\right] ,  \label{eqz21}
\end{eqnarray}%
from which the angle $\theta _{p}$ may be found by taking the
inverse sine
and cosine. The analog of Eq.\ (\ref{eq43}) is%
\begin{equation}
E_{tot}\epsilon ^{\prime }-k^{\prime }p_{cross}^{\shortparallel }=m_{e}^{2}+%
\widehat{\xi },  \label{eqz25}
\end{equation}%
where for convenience we have defined%
\begin{eqnarray}
p_{cross}^{\shortparallel } &\equiv &k\cos \theta _{e}-p\cos \left(
\theta
_{e}+\chi \right)   \label{eqz26} \\
p_{cross}^{\perp } &\equiv &k\sin \theta _{e}-p\sin \left( \theta
_{e}+\chi \right) .  \label{eqz27}
\end{eqnarray}%
One then has that
\begin{equation}
k^{\prime }=\frac{1}{\widehat{a}}\left[ \widehat{b}+E_{tot}\sqrt{\widehat{%
\xi }^{2}-m_{e}^{2}(m_{p}^{2}+\left[ p_{cross}^{\perp }\right]
^{2})}\right] \label{eqz28}
\end{equation}%
with
\begin{eqnarray}
\widehat{a} &\equiv &E_{tot}^{2}-\left[ p_{cross}^{\shortparallel
}\right]
^{2}=s+\left[ p_{cross}^{\perp }\right] ^{2}  \label{eqz29} \\
&=&m_{p}^{2}+m_{e}^{2}+2\widehat{\xi }+\left[ p_{cross}^{\perp
}\right] ^{2}
\label{eqz30} \\
\widehat{b} &\equiv &(m_{e}^{2}+\widehat{\xi
})p_{cross}^{\shortparallel }. \label{eqz31}
\end{eqnarray}%
The rest of the developments go through as before in the collinear case.


\section{Leptonic and Hadronic tensors\label{sec:tensors}}

For the leptonic (here electron) tensor one has%
\begin{equation}
\eta _{\mu \nu }=\frac{1}{2}\left[ \eta _{\mu \nu }^{unpol}+\eta
_{\mu \nu }^{pol}\right] ,  \label{eq60}
\end{equation}%
where, following standard developments \cite{Don86}, the unpolarized
tensor (symmetric under $\mu \leftrightarrow \nu $) is given by
\begin{equation}
2m_{e}^{2}\eta _{\mu \nu }^{unpol}=\frac{1}{2}Q^{2}\left( g_{\mu \nu }-\frac{%
Q_{\mu }Q_{\nu }}{Q^{2}}\right) +2R_{\mu }R_{\nu }  \label{eq61}
\end{equation}%
with
\begin{equation}
R_{\mu }\equiv \frac{1}{2}\left( K_{\mu }+K_{\mu }^{\prime }\right)
. \label{eq62}
\end{equation}%
The polarized tensor (antisymmetric under $\mu \leftrightarrow \nu
$) is given by
\begin{equation}
2m_{e}^{2}\eta _{\mu \nu }^{pol}=i\epsilon _{\mu \nu \alpha \beta
}(m_{e}S_{e}^{\alpha })Q^{\beta },  \label{eq63}
\end{equation}%
where it can be shown \cite{Don86} that the general spin 4-vector is given by%
\begin{equation}
m_{e}S_{e}^{\alpha }=h_{e}\epsilon \left( \beta _{e}\cos \mu
_{e},\cos \mu _{e}\mathbf{u}_{L}^{e}+\frac{1}{\gamma _{e}}\sin \mu
_{e}\mathbf{u}_{\bot }^{e}\right) .  \label{eq64}
\end{equation}%
Here $\mathbf{u}_{L}^{e}$ is a unit vector pointing along $\mathbf{k}$ and $%
\mathbf{u}_{\bot }^{e}$ is transverse to this direction. As usual, one has $%
\beta _{e}=k/\epsilon $ and $\gamma _{e}=\epsilon /m_{e}=[1-\beta
_{e}^{2}]^{-1/2}.$ Also, the factor $h_{e}=\pm 1$ is introduced
simply to make it easy to switch the electron's polarization from
along the beam direction to opposite to it. From this equation one
sees that transverse polarizations are suppressed by the
relativistic $\gamma $-factor and so henceforth we consider only
longitudinally polarized incident electrons:
\begin{equation}
m_{e}S_{e,L}^{\alpha }=h_{e}\epsilon (\beta
_{e},\mathbf{u}_{L}^{e}). \label{eq65}
\end{equation}%
This yields only three distinct cases for the polarized electron
tensor. Since the tensor is antisymmetric under $\mu
\leftrightarrow \nu $ we can restrict our attention to cases where
$\mu <\nu $, the others being given by using the antisymmetry. The
nonzero cases are then:
\begin{equation}
2m_{e}^{2}\eta _{\mu \nu }^{pol}=-ih_{e}z_{\mu \nu },
\label{eq65aa}
\end{equation}%
where
\begin{equation}
z_{\mu \nu }\equiv \left\{
\begin{array}{cc}
K\cdot Q=\epsilon \omega -kq\cos \theta _{q}=\frac{1}{2}Q^{2} & \mu
\nu =12
\\
\epsilon q\sin \theta _{q}=\epsilon k^{\prime }\sin \theta _{e} &
\mu \nu =02
\\
kq\sin \theta _{q}=kk^{\prime }\sin \theta _{e} & \mu \nu =23%
\end{array}%
\right.  \label{eq65a}
\end{equation}%
One can verify that $Q^{\mu }z_{\mu \nu }=0$, as should be the case.


\begin{figure}[hbt]
  \includegraphics[width=0.47\textwidth,viewport=20 151 833 498]{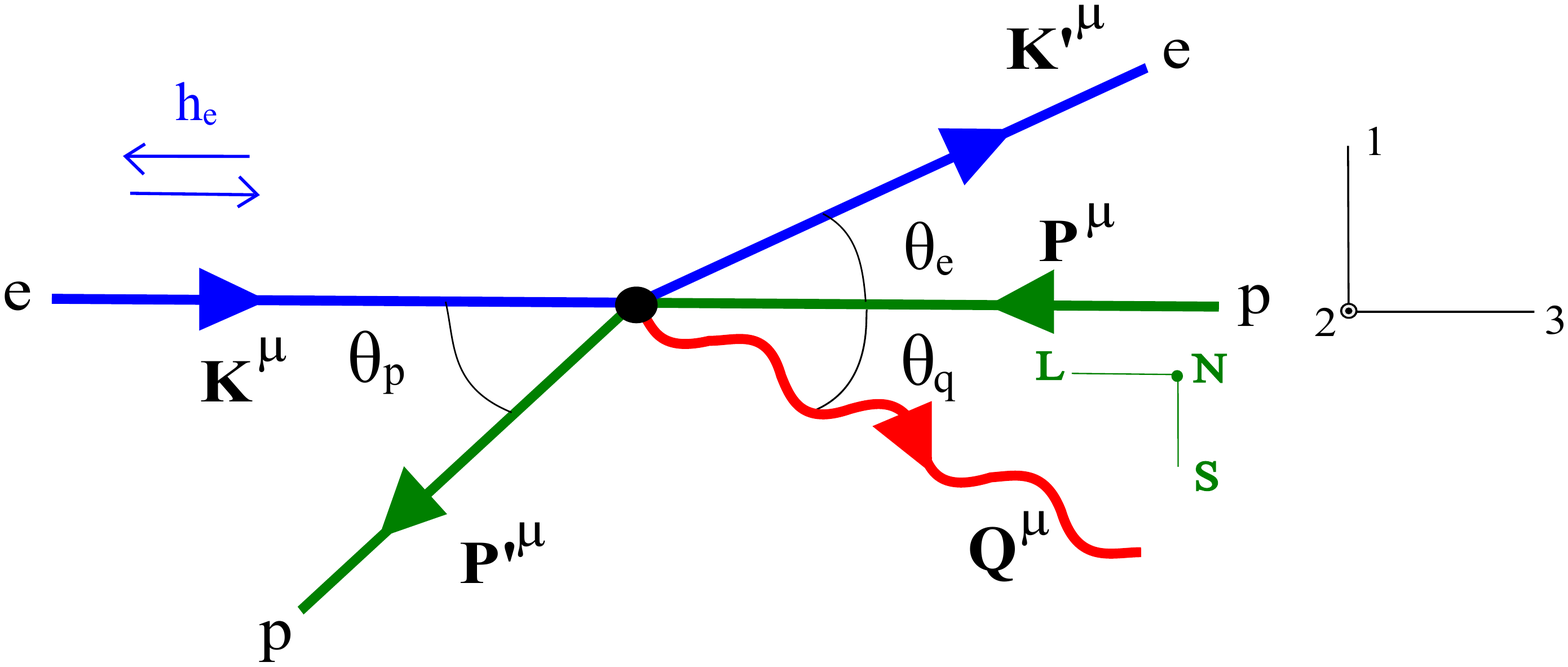}
  \caption{(color online) The electron polarization is assumed to be along the 3-axis
while the proton may be polarized longitudinally (L: along the
negative 3-axis) or sideways (S: along the negative 1-axis), as
shown.}
  \label{fig:fig3}
\end{figure}


Again restricting our attention to collinear beams at first, the
proton's
polarization 4-vector is similar to the one for the electron in Eq.\ (\ref%
{eq64}), namely
\begin{equation}
m_{p}S_{p}^{\alpha }=h_{p}E\left( \beta _{p}\cos \mu _{p},\cos \mu _{p}%
\mathbf{u}_{L}^{p}+\frac{1}{\gamma _{p}}\sin \mu
_{p}\mathbf{u}_{\bot }^{p}\right) .  \label{eq67b}
\end{equation}%
In the case of the proton we must remember that it is moving to the
left and thus, with L for longitudinal, S for sideways and N for
normal, one has
\begin{eqnarray}
\mathbf{u}_{L}^{p} &=&-\mathbf{u}_{3}  \label{eq67c} \\
\mathbf{u}_{\bot }^{p} &=&\cos \eta _{p}\mathbf{u}_{S}^{p}+\sin \eta _{p}%
\mathbf{u}_{N}^{p},  \label{eq67d}
\end{eqnarray}%
where $\mathbf{u}_{S}^{p}=-\mathbf{u}_{1}$ and $\mathbf{u}_{N}^{p}=\mathbf{u}%
_{2}$; see Fig.\ \ref{fig:fig3}. We shall specify the proton
polarization by choosing it to be along the L, S and N directions
and then any general case (for instance, when treating crossed
beams; see below) can be decomposed into components along these
three orthogonal directions. It is worthwhile to reiterate that the
conventions used here have +L polarization when it points in the
$-\mathbf{u}_{3}$ direction, +S polarization when it points in the
$-\mathbf{u}_{1}$ direction and +N polarization when it points in
the $+\mathbf{u}_{2}$ direction. One finds that
\begin{eqnarray}
\left[ S_{p}^{\alpha }\right] _{L} &=&h_{p}\gamma _{p}\left( \beta _{p},-%
\mathbf{u}_{3}\right)  \label{eq67g} \\
\left[ S_{p}^{\alpha }\right] _{S} &=&h_{p}\left(
0,-\mathbf{u}_{1}\right)
\label{eq67h} \\
\left[ S_{p}^{\alpha }\right] _{N} &=&h_{p}\left(
0,-\mathbf{u}_{2}\right) . \label{eq67i}
\end{eqnarray}

Now, for the hadronic (here proton) tensor one has the analogs of
the leptonic tensors:
\begin{equation}
W^{\mu \nu }=\frac{1}{2}\left[ W_{unpol}^{\mu \nu }+W_{pol}^{\mu \nu
}\right] ,  \label{eq68}
\end{equation}%
where the symmetric unpolarized tensor is given by
\begin{equation}
W_{unpol}^{\mu \nu }=-W_{1}\left( g^{\mu \nu }-\frac{Q^{\mu }Q^{\nu }}{Q^{2}}%
\right) +\frac{1}{m_{p}^{2}}W_{2}T^{\mu }T^{\nu }  \label{eq69}
\end{equation}%
with
\begin{equation}
T^{\mu }\equiv \frac{1}{2}\left( P^{\prime \mu }+P^{\mu }\right) .
\label{eq70}
\end{equation}%
The invariant functions $W_{1,2}$ (functions only of $Q^{2}$) are
given as
usual in terms of the proton's electromagnetic Sachs form factors by%
\begin{eqnarray}
W_{1} &=&\tau \left( G_{M}^{p}\right) ^{2}  \label{eq70a} \\
W_{2} &=&\frac{1}{1+\tau }\left[ \left( G_{E}^{p}\right) ^{2}+\tau
\left( G_{M}^{p}\right) ^{2}\right] ,  \label{eq70b}
\end{eqnarray}%
where $\tau \equiv |Q^{2}|/4m_{p}^{2}\geq 0$. The antisymmetric
polarized proton tensor is given by
\begin{eqnarray}
W_{pol}^{\mu \nu } &=&-\frac{i}{m_{p}}G_{M}^{p}\left[
G_{M}^{p}\epsilon ^{\mu \nu \alpha ^{\prime }\beta ^{\prime
}}S_{\alpha ^{\prime
}}^{p}Q_{\beta ^{\prime }}\right.  \label{eq71a} \\
&+&\left.\frac{F_{2}^{p}}{m_{p}^{2}}\left( \epsilon ^{\mu \alpha
^{\prime }\beta ^{\prime }\gamma ^{\prime }}T^{\nu }-T^{\mu
}\epsilon ^{\nu \alpha ^{\prime }\beta ^{\prime }\gamma ^{\prime
}}\right) S_{\alpha ^{\prime
}}^{p}T_{\beta ^{\prime }}Q_{\gamma ^{\prime }}\right]  \label{eq71} \nonumber \\
&\equiv &-ih_{p}\gamma _{p}\frac{1}{m_{p}}G_{M}^{p}Z^{\mu \nu }
\nonumber \\
\end{eqnarray}%
with
\begin{eqnarray}
Z^{\mu \nu } &\equiv &G_{M}^{p}Z_{1}^{\mu \nu }+\frac{1}{m_{p}^{2}}%
F_{2}^{p}Z_{2}^{\mu \nu }  \label{eq71b} \\
Z_{1}^{\mu \nu } &\equiv &\frac{h_{p}}{\gamma _{p}}\epsilon ^{\mu
\nu \alpha ^{\prime }\beta ^{\prime }}S_{\alpha ^{\prime
}}^{p}Q_{\beta ^{\prime }}
\label{eq71c} \\
Z_{2}^{\mu \nu } &\equiv &\frac{h_{p}}{\gamma _{p}}\left( \epsilon
^{\mu \alpha ^{\prime }\beta ^{\prime }\gamma ^{\prime }}T^{\nu
}-T^{\mu }\epsilon ^{\nu \alpha ^{\prime }\beta ^{\prime }\gamma
^{\prime }}\right) S_{\alpha ^{\prime }}^{p}T_{\beta ^{\prime
}}Q_{\gamma ^{\prime }}.  \label{eq71d}
\end{eqnarray}%
One can verify that $Q_{\mu }Z^{\mu \nu }=0$ as should be the case;
in fact, $Q_{\mu }Z_{1}^{\mu \nu }=Q_{\mu }Z_{2}^{\mu \nu }=0$. For
reasons that will become clear in the following section, in these
expressions it proves useful to use a mixture of the Sachs magnetic
form factor with the Pauli form factor; the Sachs and Dirac/Pauli
form factors are related in the familiar
way:%
\begin{eqnarray}
G_{E}^{p} &=&F_{1}^{p}-\tau
F_{2}^{p}\;\;\;\;\;\;\;\;\;\;\;\;\;\;\;\;\;\;\;G_{M}^{p}=F_{1}^{p}+F_{2}^{p}
\label{eq90a} \\
F_{1}^{p} &=&\frac{1}{1+\tau }\left[ G_{E}^{p}+\tau G_{M}^{p}\right]
\;\;\;\;F_{2}^{p}=\frac{1}{1+\tau }\left[ G_{M}^{p}-G_{E}^{p}\right]
\nonumber
\end{eqnarray}


\section{Contractions of Leptonic and Hadronic Tensors\label{sec:contract}}

We start by obtaining the contraction of the two symmetric
unpolarized tensors, namely
\begin{eqnarray}
X^{unpol} &\equiv &\left\{ 2m_{e}^{2}\eta _{\mu \nu
}^{unpol}\right\} \times
\left\{ W_{unpol}^{\mu \nu }\right\}  \label{eq72} \\
&=&-W_{1}\left( \frac{3}{2}Q^{2}+2R^{2}\right)\label{eq74} \\
&& \qquad+\frac{1}{m_{p}^{2}}%
W_{2}\left( \frac{1}{2}Q^{2}T^{2}+2(R\cdot T)^{2}\right) .
\nonumber
\end{eqnarray}%
One has $\frac{3}{2}Q^{2}+2R^{2}=Q^{2}+2m_{e}^{2}$ and $%
T^{2}=m_{p}^{2}-Q^{2}/4$ and, using the fact that
\begin{equation}
R\cdot T=K\cdot P+Q^{2}/4=\xi +Q^{2}/4,  \label{eq74b}
\end{equation}%
one therefore has
\begin{equation}
\frac{1}{2}Q^{2}T^{2}+2(R\cdot T)^{2}=2\left[ \xi ^{2}+\frac{1}{2}Q^{2}\xi +%
\frac{1}{4}m_{p}^{2}Q^{2}\right] .  \label{eq74c}
\end{equation}%
Defining dimensionless variables $\lambda \equiv \omega /2m_{p}$ and
$\kappa \equiv q/2m_{p}$, where then $\tau =\kappa ^{2}-\lambda
^{2},$ as usual, and
defining%
\begin{equation}
\widetilde{\epsilon }\equiv \xi /m_{p}^{2},  \label{eq74cx}
\end{equation}%
one then has
\begin{eqnarray}
X^{unpol}&\equiv& -W_{1}\left( Q^{2}+2m_{e}^{2}\right)\nonumber\\
&&\qquad+2m_{p}^{2}W_{2}\left[ \widetilde{\epsilon }^{2}-2\tau
\widetilde{\epsilon }-\tau \right] \label{eq74ff}
\end{eqnarray}%
To connect with standard notation let us use the following \footnote{%
Note that the angle $\theta _{e}^{\prime }$ defined via Eq.
(\ref{eq74hh}) is not the true scattering angle, but is an effective
angle that makes the expressions to follow relatively compact. When
one goes to the proton rest system and invokes the electron extreme
relativistic limit this primed angle reverts to the true scattering
angle $\theta_e$.}
\begin{eqnarray}
V_{0} &\equiv &4m_{p}^{2}\left[ \widetilde{\epsilon }^{2}-2\tau \widetilde{%
\epsilon }-\tau \right]  \label{eq74gg} \\
&=&4m_{p}^{2}\Big[ \left( \frac{1}{2m_{p}^{2}}\left\{ (\epsilon
+\epsilon ^{\prime })E+(k+k^{\prime }\cos \theta _{e})p\right\}
\right) ^{2} \nonumber\\
&&\qquad \qquad\qquad\qquad\qquad -\tau(1+\tau)\Big]   \label{eq74gg-1} 
\end{eqnarray}
\begin{eqnarray} 
\tan ^{2}\theta _{e}^{\prime }/2 &\equiv &\frac{-Q^{2}}{V_{0}}=\frac{\tau }{%
\widetilde{\epsilon }^{2}-2\tau \widetilde{\epsilon }-\tau }
\label{eq74hh}
\end{eqnarray}
and then
\begin{equation}
X^{unpol}=\frac{2m_{p}^{2}\tau }{\tan ^{2}\theta _{e}^{\prime
}/2}F^{2}(\tau ,\theta _{e})  \label{eq74-6}
\end{equation}%
where, defining
\begin{equation}
\mathcal{E}^{\prime }\equiv \left[ 1+2(1+\tau )\tan ^{2}\theta
_{e}^{\prime }/2\left( 1+\frac{2m_{e}^{2}}{Q^{2}}\right) \right]
^{-1},  \label{eq74-6a}
\end{equation}%
one has for the total (squared) e-p scattering form factor
\begin{eqnarray}
F^{2}(\tau ,\theta _{e}) &=&W_{2}+2W_{1}\tan ^{2}\theta _{e}^{\prime
}/2\left( 1+\frac{2m_{e}^{2}}{Q^{2}}\right)  \label{eq74k} \\
&=&\frac{1}{(1+\tau )\mathcal{E}^{\prime }}\left[
\mathcal{E}^{\prime }\left( G_{E}^{p}\right) ^{2}+\tau \left(
G_{M}^{p}\right) ^{2}\right] . \label{eq74kz}
\end{eqnarray}

It is useful at this point to check this general result for the
special case of the laboratory frame; there $p=0$ and so $E=m_{p}$;
also $\lambda
_{lab}=\tau $ and $\kappa _{lab}=\sqrt{\tau (1+\tau )}$. This implies that $%
\xi _{lab}=m_{p}\epsilon $ and thus $\widetilde{\epsilon
}_{lab}=\epsilon
/m_{p}$ and then in the lab.\ system one can show that $V_{0}^{lab}=v_{0}=(%
\epsilon +\epsilon ^{\prime })^{2}-q^{2}$, the usual answer
\cite{Don86}, and one then has
\begin{equation}
X_{lab}^{unpol}\equiv \frac{1}{2}v_{0}F^{2}(\tau ,\theta _{e}).
\label{eq74j}
\end{equation}%
Furthermore, in the electron Extreme Relativistic Limit (ERL$_{e}$)
where the electron's mass may be neglected with respect to its
momentum, $\theta _{e}^{\prime }\rightarrow \theta _{e}$,
$\mathcal{E}^{\prime }\rightarrow \mathcal{E}$ (the usual so-called
virtual photon longitudinal polarization) and the expression above
becomes proportional to $\mathcal{E}\left( G_{E}^{p}\right)
^{2}+\tau \left( G_{M}^{p}\right) ^{2}$, the familiar answer.

For the contraction of the two antisymmetric tensors we have from
the expressions above
\begin{eqnarray}
X^{pol} &\equiv &\left\{ 2m_{e}^{2}\eta _{\mu \nu }^{pol}\right\}
\times
\left\{ W_{pol}^{\mu \nu }\right\}  \label{eq76} \\
&=&-h_{e}h_{p}\gamma _{p}\frac{1}{m_{p}}G_{M}^{p}z_{\mu \nu }Z^{\mu
\nu }, \label{eq77}
\end{eqnarray}%
where $Z^{\mu \nu }$ may be decomposed into $Z_{1}^{\mu \nu }$ and $%
Z_{2}^{\mu \nu }$ as in Eq.\ (\ref{eq71b}). Since the choice of
longitudinal electron polarization led to only the components $\mu
\nu =12,$ 02 and 23 (together with their reverses, which, using the
antisymmetry, leads to an overall factor of 2 if only this order is
retained), we have only three cases to consider. Furthermore, note
that all cases here have either $\mu $ or $\nu $ equal to 2 and so
the proton polarization cannot have component 2. Since this is the
only component for N polarization (see Eq.\ (\ref{eq67i})) we find
that the proton's polarization (in one-photon-exchange
approximation) \emph{cannot be normal}, as expected. For the tensors
of type
1 the results are the following:%
\begin{equation}
\left[ Z_{1}^{\mu \nu }\right] _{L}=\left\{
\begin{array}{ll}
\omega +\beta _{p}q\cos \theta _{q}=-Q^{2}/2E & \mu \nu =12 \\
-q\sin \theta _{q} & \mu \nu =02 \\
-\beta _{p}q\sin \theta _{q} & \mu \nu =23%
\end{array}%
\right.  \label{eq80a}
\end{equation}%
and
\begin{equation}
\left[ Z_{1}^{\mu \nu }\right] _{S}=\frac{1}{\gamma _{p}}\left\{
\begin{array}{ll}
0 & \mu \nu =12 \\
-q\cos \theta _{q} & \mu \nu =02 \\
\omega & \mu \nu =23%
\end{array}%
\right.  \label{eq80b}
\end{equation}%
with no allowed N components. For the tensors of type 2 one has%
\[
h_{p}\gamma _{p}Z_{2}^{2\mu }=T^{\mu }Z_{0},
\]%
with $\mu =0,1,$ or 3 and
\begin{equation}
Z_{0}\equiv \epsilon ^{2\alpha ^{\prime }\beta ^{\prime }\gamma
^{\prime }}S_{\alpha ^{\prime }}^{p}T_{\beta ^{\prime }}Q_{\gamma
^{\prime }} \label{eq79}
\end{equation}%
yielding
\begin{eqnarray}
h_{p}\left[ Z_{0}\right] _{L} &=&-m_{p}q\sin \theta _{q}  \label{eq81aa} \\
h_{p}\left[ Z_{0}\right] _{S} &=&-\left[ p\omega +Eq\cos \theta
_{q}\right]
\label{eq81bb} \\
\left[ Z_{0}\right] _{N} &=&0.  \label{eq81b}
\end{eqnarray}

The polarized contraction in Eq.\ (\ref{eq77}) is then given by
\begin{equation}
X^{pol}=-h_{e}h_{p}4EG_{M}^{p}C^{pol},  \label{eq82}
\end{equation}%
where
\begin{eqnarray}
C^{pol} &\equiv &\frac{1}{8m_{p}^{3}}z_{\mu \nu }Z^{\mu \nu }\equiv
G_{M}^{p}C_{1}^{pol}+F_{2}^{p}C_{2}^{pol}  \label{eq82a} \\
&=&\frac{1}{4m_{p}^{3}}\left(
z_{12}Z^{12}+z_{02}Z^{02}+z_{23}Z^{23}\right) \label{eq83b}
\end{eqnarray}%
with $z_{\mu \nu }$ from Eq.\ (\ref{eq65a}) and $Z^{\mu \nu }$ from
the developments given above, with the subscripts 1 and 2 referring
to the two contributions in $Z^{\mu \nu }$. The extra factor of 2 in
Eq.\ (\ref{eq83b}) comes from using the antisymmetry together with
only one order of the indices $\mu \nu $. Again employing
dimensionless variables one has
\begin{eqnarray}
C_{1L}^{pol} &=&-\frac{1}{\gamma _{p}}\left[ \tau ^{2}+\widetilde{\epsilon }%
\kappa ^{2}\sin ^{2}\theta _{q}\right]  \label{eq88a} \\
C_{1S}^{pol} &=&\kappa \sin \theta _{q}\Big\{ (\lambda +\beta
_{p}\kappa \cos \theta _{q})\sqrt{\widetilde{\epsilon
}^{2}-(m_{e}/m_{p})^{2}}
\nonumber \\
&&\qquad\qquad -\left( \beta _{p}\lambda +\kappa \cos \theta
_{q}\Big\}
\widetilde{\epsilon }\right\}  \label{eq88bb} \\
C_{2L}^{pol} &=&\frac{1}{\gamma _{p}}\widetilde{\epsilon }\kappa
^{2}\sin
^{2}\theta _{q}  \label{eq88c} \\
C_{2S}^{pol} &=&\widetilde{\epsilon }\kappa \sin \theta _{q}\left(
\beta _{p}\lambda +\kappa \cos \theta _{q}\right) ,  \label{eq88d}
\end{eqnarray}%
where the result in Eq.\ (\ref{eq88bb}) is obtained using the facts
that
\begin{eqnarray}
k &=&E\sqrt{\widetilde{\epsilon }^{2}-(m_{e}/m_{p})^{2}}-p\widetilde{%
\epsilon }  \label{eq89} \\
\epsilon &=&E\widetilde{\epsilon }-p\sqrt{\widetilde{\epsilon }%
^{2}-(m_{e}/m_{p})^{2}}.  \label{eq90}
\end{eqnarray}

It can be shown that $C_{1L}^{pol}$, $C_{2L}^{pol}$ and
$C_{2S}^{pol}$ are
all of order unity when $\gamma _{p}\rightarrow \infty $, whereas $%
C_{1S}^{pol}$ goes as $1/\gamma _{p}$ in that limit. Furthermore, if
one sets $F_{1}^{p}=1$ and $F_{2}^{p}=0,$ then, through Eqs.\ (\ref{eq90a}), one has $G_{E}^{p}=G_{M}^{p}=1$, and thus no terms of
type 2 contribute (no terms involving $C_{2L}^{pol}$ or
$C_{2S}^{pol}$). This special case makes the proton current take on
the same form as that of a point Dirac particle like the electron.
Accordingly one sees that the only surviving contribution in the
extreme relativistic limit for collisions of point Dirac particles
is the one involving $C_{1L}^{pol}$, as expected. This also
underlines the fact that the sideways contribution in the general
case for ultra-relativistic protons arises because of the anomalous
magnetic moment, \textit{i.e.,} the parts of the current involving
$F_{2}$.

Finally, using the relationship for the Pauli form factor in terms
of Sachs form factors (Eqs.\ (\ref{eq90a})) and defining
\begin{eqnarray}
C_{M}^{pol} &\equiv &\frac{1}{1+\tau }\left[ (1+\tau )C_{1}^{pol}+C_{2}^{pol}%
\right]  \label{eq102} \\
C_{E}^{pol} &\equiv &\frac{1}{1+\tau }\left[ -C_{2}^{pol}\right] ,
\label{eq103}
\end{eqnarray}%
then one can write%
\begin{equation}
C^{pol}=C_{M}^{pol}G_{M}^{p}+C_{E}^{pol}G_{E}^{p}.  \label{eq104}
\end{equation}%
From above we therefore have that
\begin{eqnarray}
C_{ML}^{pol} &=&-\frac{1}{\gamma _{p}}\frac{\tau }{1+\tau }\left[
\tau (1+\tau )+\widetilde{\epsilon }\kappa ^{2}\sin ^{2}\theta
_{q}\right]
\label{eq105} \\
C_{MS}^{pol} &=&\frac{1}{1+\tau }\kappa \sin \theta _{q}\Big\{
(1+\tau
)(\lambda +\beta _{p}\kappa \cos \theta _{q})  \nonumber \\
&&\qquad\qquad\qquad\times \sqrt{\widetilde{\epsilon }%
^{2}-(m_{e}/m_{p})^{2}}  \nonumber \\
&&\qquad\qquad\qquad-\tau \left( \beta _{p}\lambda +\kappa \cos \theta
_{q}\right)
\widetilde{\epsilon }\Big\}  \label{eq106a} \\
C_{EL}^{pol} &=&-\frac{1}{\gamma _{p}}\frac{1}{1+\tau }\widetilde{\epsilon }%
\kappa ^{2}\sin ^{2}\theta _{q}  \label{eq107} \\
C_{ES}^{pol} &=&-\frac{1}{1+\tau }\widetilde{\epsilon }(\beta
_{p}\lambda +\kappa \cos \theta _{q})\kappa \sin \theta _{q}.
\label{eq108}
\end{eqnarray}


\begin{figure}[hbt]
  \includegraphics[width=0.47\textwidth,viewport=24 175 833 493]{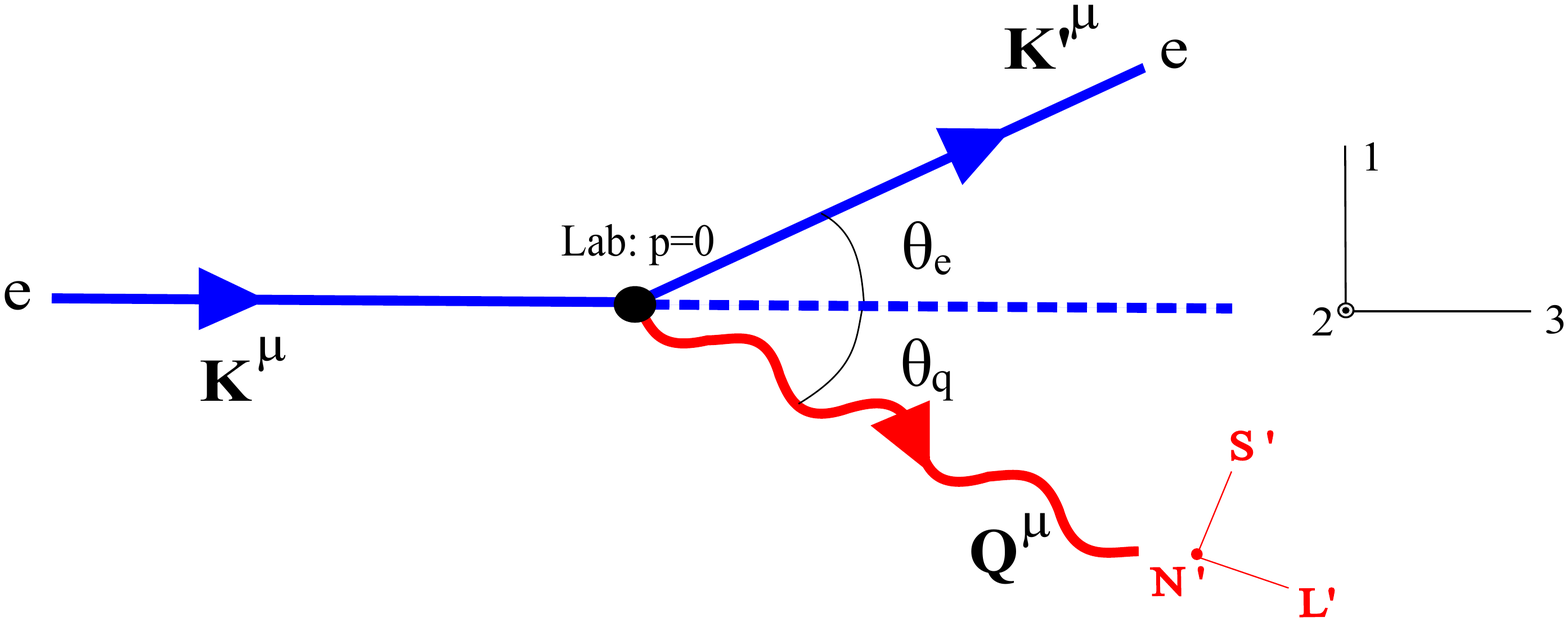}
  \caption{(color online) Proton polarizations $(L^{\prime },S^{\prime },N^{\prime })$
in the lab.\ system where $p=0$.}
  \label{fig:fig4}
\end{figure}


Next we will want to check these results by going to the lab.\ frame where $%
p=0,$ $E=m_{p},$ $\beta _{p}=0,$ $\gamma _{p}=1,$ $\widetilde{\epsilon }%
=\epsilon /m_{p},$ $\sqrt{\widetilde{\epsilon }^{2}-(m_{e}/m_{p})^{2}}%
=k/m_{p},\lambda =\tau $ and $\kappa =\sqrt{\tau (1+\tau )}$. One
has
\begin{eqnarray}
\left[ C_{ML}^{pol}\right] ^{lab} &=&-\tau ^{2}\left( 1+\frac{\epsilon }{%
m_{p}}\sin ^{2}\theta _{q}\right)  \label{eq91a} \\
\left[ C_{MS}^{pol}\right] ^{lab} &=&\tau\frac{1}{m_{p}}\Big[\sqrt{\tau (1+\tau )}k \nonumber\\
&&\qquad\qquad\qquad-\tau \epsilon \cos \theta _{q}\Big]\sin\theta _{q}
\label{eq91b} \\
\left[ C_{EL}^{pol}\right] ^{lab} &=&-\tau \frac{\epsilon
}{m_{p}}\sin
^{2}\theta _{q}  \label{eq91c} \\
\left[ C_{ES}^{pol}\right] ^{lab} &=&-\tau \frac{\epsilon
}{m_{p}}\sin \theta _{q}\cos \theta _{q}.  \label{eq91d}
\end{eqnarray}%
In the lab.\ system the polarizations of the proton are usually
specified with respect to the $(L^{\prime },S^{\prime },N^{\prime
})$ coordinate system as shown in Fig.\ \ref{fig:fig4}. Rotating to
this system one has
\begin{eqnarray}
\left[ C_{XL^{\prime }}^{pol}\right] ^{lab} &=&-\cos \theta
_{q}\left[ C_{XL}^{pol}\right] ^{lab}+\sin \theta _{q}\left[
C_{XS}^{pol}\right] ^{lab}
\label{eq109} \\
\left[ C_{XS^{\prime }}^{pol}\right] ^{lab} &=&-\sin \theta
_{q}\left[ C_{XL}^{pol}\right] ^{lab}-\cos \theta _{q}\left[
C_{XS}^{pol}\right] ^{lab} \label{eq110}
\end{eqnarray}%
with $X=M$ or $E$. After some algebra one finds that
\begin{eqnarray}
\left[ C_{ML^{\prime }}^{pol}\right] ^{lab} &=&\tau \Big[ \tau \cos
\theta _{q}\nonumber \\
&& \qquad+\sqrt{\tau (1+\tau )}\frac{k}{m_{p}}\sin ^{2}\theta
_{q}\Big]
\label{eq115} \\
\left[ C_{MS^{\prime }}^{pol}\right] ^{lab} &=&0  \label{eq116} \\
\left[ C_{EL^{\prime }}^{pol}\right] ^{lab} &=&0  \label{eq117} \\
\left[ C_{ES^{\prime }}^{pol}\right] ^{lab} &=&\frac{\epsilon
}{m_{p}}\tau \sin \theta _{q},  \label{eq118}
\end{eqnarray}%
where the zeros in Eqs.\ (\ref{eq116}) and (\ref{eq117}) are expected
from the familiar lab.\ frame analysis.


\section{The Cross Section and Polarization Asymmetry\label{sec:cross-asym}}

Finally in these developments of the formalism, we now want to
obtain the cross section and polarization asymmetry in the collider
frame, together with their lab.\ frame and extreme relativistic
limits. We begin with the unpolarized cross section in the general
collider frame. First, the flux factor to be used in applying the
Feynman rules must now be generalized (see \cite{BjD64}, Eqs.(\ 7.41)
and (B.1): where in the lab.\ frame one has the multiplicative factor
$1/\beta _{e}\gamma _{e}$ one now has the replacement
\begin{equation}
\frac{1}{\beta _{e}\gamma _{e}}\rightarrow \frac{1}{\gamma
_{e}\gamma _{p}\left( \beta _{e}+\beta _{p}\right) },  \label{eq119}
\end{equation}%
where both factors of $\beta $ are to be taken positive. Clearly the
lab.\ frame result emerges when $\beta _{p}\rightarrow 0$ and $\gamma
_{p}\rightarrow 1$. The recoil factor can be shown to generalize to
\begin{equation}
F_{rec}=1+\frac{\epsilon k^{\prime }-\epsilon ^{\prime }(k-p)\cos \theta _{e}%
}{Ek^{\prime }}  \label{eq120}
\end{equation}%
and using $V_{0}$ given in Eq.\ (\ref{eq74gg}) one has
\begin{equation}
\left[ \frac{d\sigma }{d\Omega _{e}}\right] _{ep}^{unpol,\text{\textit{%
collider}}}=\sigma _{M}^{\text{\textit{collider}}}\left(
F_{rec}\right) ^{-1}F^{2}(\tau ,\theta _{e})  \label{eq122}
\end{equation}%
with the square of the form factor from Eq.\ (\ref{eq74k}) and the
generalized Mott cross section given by
\begin{equation}
\sigma _{M}^{\text{\textit{collider}}}=\left( \frac{\alpha
}{Q^{2}}\right)
^{2}\frac{k^{\prime }}{k}V_{0}\frac{\beta _{e}}{\beta _{e}+\beta _{p}}\frac{1%
}{\gamma _{p}^{2}}.  \label{eq123}
\end{equation}%
These results can be checked by going to the lab.\ frame and shown to
agree with the familiar answers.

Various limiting cases may be straightforwardly obtained. First, in the ERL$%
_{e}$ one has
\begin{eqnarray}
F_{rec}^{\text{\textit{ERL}}_{e}} &=&1+\beta _{p}\cos \theta _{e}
\label{eq128} \\
\widetilde{\epsilon }^{\text{\textit{ERL}}_{e}}
&=&\frac{k}{m_{p}}\gamma
_{p}1+\beta _{p}  \label{eq128a} \\
\sigma _{M}^{\text{\textit{collider,ERL}}_{e}} &=&\left( \frac{\alpha }{Q^{2}%
}\right) ^{2}\frac{k^{\prime }}{k}V_{0}^{\text{\textit{ERL}}_{e}}\frac{1}{%
1+\beta _{p}}\frac{1}{\gamma _{p}^{2}}  \label{eq129}
\end{eqnarray}%
where $V_{0}^{\text{\textit{ERL}}_{e}}$ may be obtained using Eq.\ (\ref%
{eq74gg}) and therefore $\sigma
_{M}^{\text{\textit{collider,ERL}}_{e}}$ using Eq.\ (\ref{eq123}).

With both beams polarized the cross section has two terms, one
($\Sigma $) containing no dependence on the polarizations and one
($\Delta $) containing
only terms where both beams are polarized \footnote{%
Here we consider only parity-conserving $e-p$ scattering; for the
parity-violating situation where only the electron beam is assumed
to be
polarized, see Sect.\ VI.}, the latter being proportional to the product $%
h_{e}h_{p}$:
\begin{equation}
\left[ \frac{d\sigma }{d\Omega }\right] _{ep}^{pol,\text{\textit{collider}}%
}\equiv \sigma ^{h_{e},h_{p}}=\Sigma +h_{e}h_{p}\Delta .
\label{eq136}
\end{equation}%
By flipping the spins one can form the polarization asymmetry:
\begin{equation}
\left[ A\right] _{ep}^{pol,\text{\textit{collider}}}\equiv
\frac{\sigma ^{+1,+1}-\sigma ^{-1,+1}}{\sigma ^{+1,+1}+\sigma
^{-1,+1}}=\frac{\sigma
^{+1,+1}-\sigma ^{+1,-1}}{\sigma ^{+1,+1}+\sigma ^{+1,-1}}=\frac{\Delta }{%
\Sigma }.  \label{eq137}
\end{equation}%
Using the developments above we immediately have that
\begin{equation}
\left[ A\right] _{ep}^{pol,\text{\textit{collider}}}=\frac{h_{e}h_{p}X^{pol}%
}{X^{unpol}},  \label{eq138}
\end{equation}%
where $X^{unpol}$ is given in Eq.\ (\ref{eq74-6}) and $X^{pol}$ is
defined in Eq.\ (\ref{eq77}). Substituting for these and expressing
$C^{pol}$ in terms of $G_{M}^{p}$ and $G_{E}^{p}$ using Eq.
(\ref{eq104}) we find that
\begin{equation}
\left[ A\right] _{ep}^{pol,\text{\textit{collider}}}=-\frac{\mathcal{N}}{%
F^{2}(\tau ,\theta _{e})},  \label{eq139}
\end{equation}%
where the numerator is given by
\begin{eqnarray}
\mathcal{N} &\equiv &\frac{2}{\tau }\gamma _{p}\tan ^{2}\theta
_{e}^{\prime
}/2\left[ G_{M}^{p}\left( C_{M}^{pol}G_{M}^{p}+C_{E}^{pol}G_{E}^{p}\right) %
\right]  \label{eq139a} \\
&\equiv &\mathcal{N}_{M}\left( G_{M}^{p}\right) ^{2}+\mathcal{N}%
_{E}G_{E}^{p}G_{M}^{p}  \label{eq139b}
\end{eqnarray}%
and where $C_{M,E}^{pol}$ are given in Eqs.\ (\ref{eq105}-\ref{eq108}) for the two types of proton polarization,
$L$ and $S$. Again, for reference, the denominator in Eq.
(\ref{eq139}) is given in Eq.\ (\ref{eq74k}) and $\tan ^{2}\theta
_{e}^{\prime }/2$ is given in Eq.\ (\ref{eq74hh}).

As a first check, let us go to the lab.\ frame and consider the
$L^{\prime }$
and $S^{\prime }$ polarizations discussed above. We have from Eq.\ (\ref%
{eq115}) that the $L^{\prime }$ part of the numerator in Eq.\ (\ref%
{eq139a}) in the lab.\ frame where $\theta _{e}^{\prime }\rightarrow
\widetilde{\theta }_{e}$ is given by
\begin{eqnarray}
\lefteqn{-\frac{2}{\tau }\tan ^{2}\widetilde{\theta }_{e}/2\left[
C_{ML^{\prime
}}^{pol}\right] ^{lab}\left( G_{M}^{p}\right) ^{2} }   \nonumber \\
&=&-2\tan ^{2}\widetilde{\theta }_{e}/2 \nonumber\\
&& \times \left[ \tau \cos \theta _{q}+\sqrt{%
\tau (1+\tau )}\frac{k}{m_{p}}\sin ^{2}\theta _{q}\right]  \left (G_{M}^{p}\right) ^{2}  \label{eq140a} \\
&\equiv &V_{T^{\prime }}W_{L^{\prime }}^{T^{\prime }},
\label{eq141}
\end{eqnarray}%
using a notation where
\begin{equation}
W_{L^{\prime }}^{T^{\prime }}=-2\tau \left( G_{M}^{p}\right) ^{2}.
\label{eq142}
\end{equation}%
This implies that
\begin{eqnarray}
V_{T^{\prime }} &=&\frac{1}{\tau }\tan ^{2}\widetilde{\theta
}_{e}/2 \nonumber\\
&&\qquad \times \left[ \tau \cos \theta _{q}+\sqrt{\tau (1+\tau
)}\frac{k}{m_{p}}\sin ^{2}\theta
_{q}\right]  \label{eq143} \\
&=&\frac{1}{\beta }\tan ^{2}\widetilde{\theta }_{e}/2\nonumber\\
&&\qquad \times \left(\frac{\epsilon +\epsilon ^{\prime }}{q}\right) \left[
1-\frac{2m_{e}^{2}q^{2}}{\epsilon (\epsilon +\epsilon ^{\prime
})(-Q^{2})}\right] ,  \label{eq144}
\end{eqnarray}%
which agrees with the standard notation (\textit{e.g.,} see
\cite{Don86}). The $S^{\prime }$ part of the numerator in Eq.
(\ref{eq139}) in the lab.\ frame is found similarly:
\begin{eqnarray}
\lefteqn{-\frac{2}{\tau }\tan ^{2}\widetilde{\theta }_{e}/2\left[C_{ES^{\prime
}}^{pol}\right] ^{lab}G_{E}^{p}G_{M}^{p} =2\tan ^{2}\widetilde{\theta }_{e}/2} \nonumber \\
&&\qquad \qquad\qquad\qquad \times \left( \frac{\epsilon }{m_{p}}\right) \sin \theta_{q}G_{E}^{p}G_{M}^{p}  \label{eq145} \\
&\equiv &V_{TL^{\prime }}W_{S^{\prime }}^{TL^{\prime }},
\label{eq146}
\end{eqnarray}%
again using the notation where
\begin{equation}
W_{S^{\prime }}^{TL^{\prime }}=2\sqrt{2\tau (1+\tau
)}G_{E}^{p}G_{M}^{p} \label{eq147}
\end{equation}%
yielding
\begin{equation}
V_{TL^{\prime }}=-\frac{1}{v_{0}}\sqrt{2}\frac{\tau }{\kappa
^{2}}\epsilon k^{\prime }\sin \theta _{e},  \label{eq148}
\end{equation}%
which agrees with the standard notation (\textit{e.g.,} see
\cite{Don86}).

Finally, for the crossed beams situation one must extend the
discussion of the leptonic and hadronic tensors in Sect.
\ref{sec:tensors}. In this case it is more convenient to work in the
($1,2,3$) system. The leptonic tensor will be as before; however,
the hadronic tensor will now be assumed to have L and S
polarizations with respect to the rotated frame. That is, we will
assume that L- or S-polarizations occur when the proton's
polarization lies along
the directions%
\begin{eqnarray}
\mathbf{u}_{L,\mathrm{crossed}} &=&-\mathbf{u}_{z^{\prime }}=\sin
\chi
\mathbf{u}_{1}-\cos \chi \mathbf{u}_{3}  \label{eqz32} \\
\mathbf{u}_{S,\mathrm{crossed}} &=&-\mathbf{u}_{x^{\prime }}=-\cos
\chi \mathbf{u}_{1}-\sin \chi \mathbf{u}_{3}.  \label{eqz33}
\end{eqnarray}%
Accordingly the asymmetries in the crossed beams situation are
simply linear
combinations of the collinear ones that we derived in Sect. \ref%
{sec:cross-asym}:%
\begin{eqnarray}
\left. \left[ A\right] _{ep}^{pol,\text{\textit{collider}}}\right\vert _{L,%
\mathrm{crossed}} &=&-\sin \chi \left. \left[ A\right] _{ep}^{pol,\text{%
\textit{collider}}}\right\vert _{S,\mathrm{collinear}}\nonumber\\
&&+\cos \chi \left. %
\left[ A\right] _{ep}^{pol,\text{\textit{collider}}}\right\vert _{L,\mathrm{%
collinear}}  \label{eqz34} \\
\left. \left[ A\right] _{ep}^{pol,\text{\textit{collider}}}\right\vert _{S,%
\mathrm{crossed}} &=&\cos \chi \left. \left[ A\right] _{ep}^{pol,\text{%
\textit{collider}}}\right\vert _{S,\mathrm{collinear}}\nonumber\\
&&+\sin \chi \left. %
\left[ A\right] _{ep}^{pol,\text{\textit{collider}}}\right\vert _{L,\mathrm{%
collinear}}.  \label{eqz35}
\end{eqnarray}

It is straightforward to verify that all of the collinear results
obtained above are recovered when the crossing angle $\chi $ is set
to zero.


\section{Parity-Violating Elastic Electron Scattering\label{sec:PV}}


\begin{figure}[hbt]
  \includegraphics[width=0.4\textwidth,viewport=80 309 491 739]{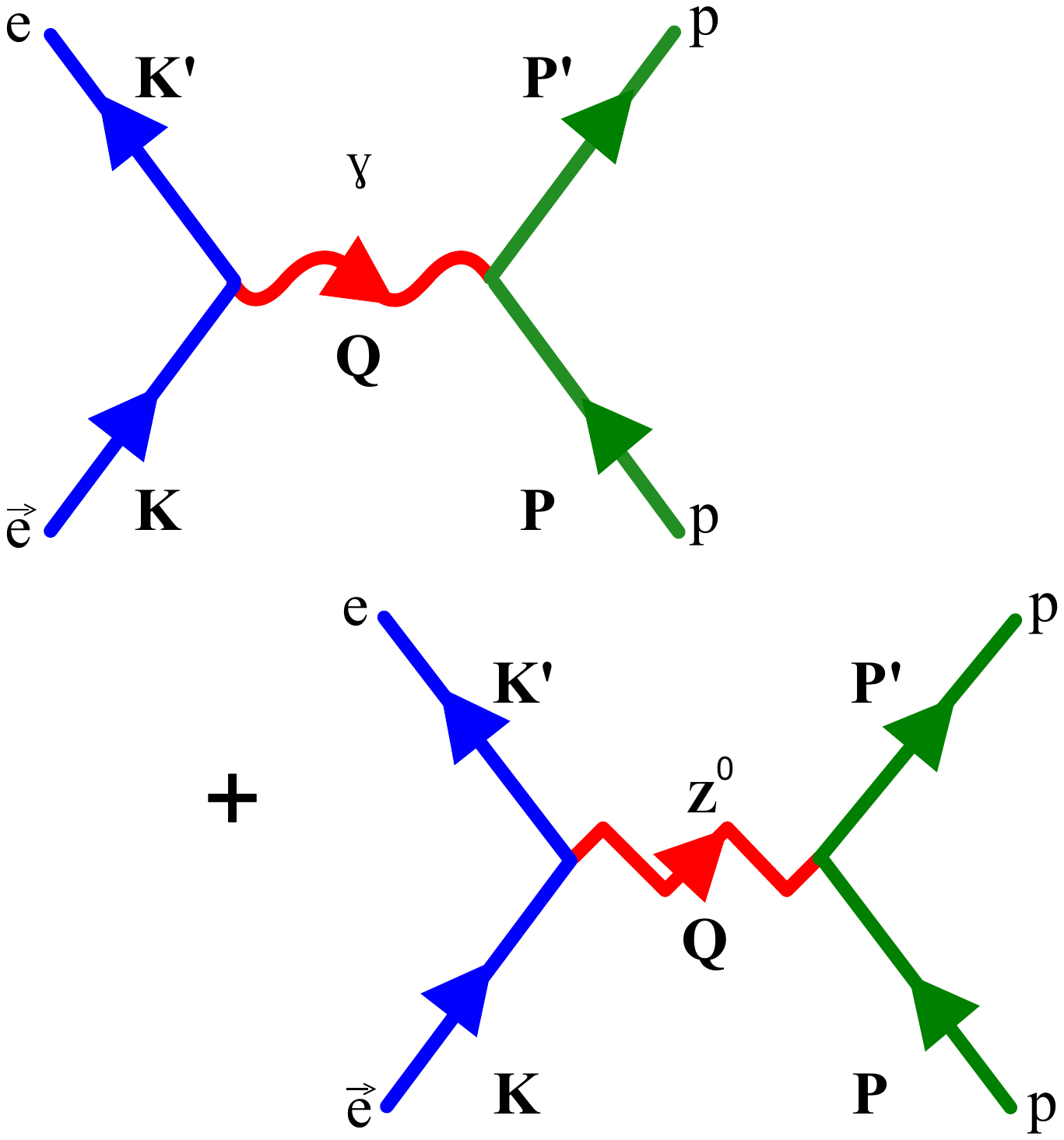}
  \caption{(color online) Parity-violating electron-proton elastic scattering.}
  \label{fig:fig5}
\end{figure}


While our main focus in the present work is placed
on the double-polarization reaction $\overrightarrow{p}(\overrightarrow{e}%
,e)p$ where parity-conserving (PC) asymmetries occur, it is
interesting also
to consider in context parity-violating (PV) single-polarization $p(%
\overrightarrow{e},e)p$ scattering. The general structure of PV
elastic electron scattering from the proton is illustrated in Fig.
\ref{fig:fig5}. Two diagrams are relevant: the usual photon exchange
diagram which is parity conserving and a $Z^{0}$ exchange diagram
which has both polar vector (V) and axial vector (A) contributions.
To obtain the total cross section one must take the sum of these two
diagrams and compute the square of the absolute value of that sum,
leading to terms from the square of the photon exchange diagram (VV)
which are parity conserving, terms from the interferences between
the two diagrams which include VA pieces that are parity violating
(PV) and pieces from the square of the $Z^{0}$ exchange diagram
which are very small and so neglected. Upon considering
longitudinally polarized electron scattering elastic from
\emph{unpolarized} protons, the electron spin asymmetry can only
occur from parity violations and thus the VA interferences are the
leading-order PV contributions. The details of the developments of
the PV formalism are given in several references \cite{DDS88,Mus94}
and here we only summarize the differences that occur when working
in the collider frame. We begin with a brief discussion of the
leptonic and hadronic tensors involved.

In these developments for simplicity we consider only the extreme
relativistic limit for the electrons. In Sect. \ref{sec:tensors} the
leptonic
tensor for the unpolarized situation was given (see Eq.\ (\ref{eq61})):%
\begin{equation}
\chi _{\mu \nu }^{(s)}\equiv \frac{1}{2}Q^{2}\left( g_{\mu \nu }-\frac{%
Q_{\mu }Q_{\nu }}{Q^{2}}\right) +2R_{\mu }R_{\nu }  \label{eqPV1}
\end{equation}%
which is symmetric under interchange of $\mu $ and $\nu $. Now,
since the weak interaction has both vector and axial vector
contributions, a second
tensor must be considered:%
\begin{equation}
\chi _{\mu \nu }^{(a)}\equiv i\epsilon _{\mu \nu \alpha \beta
}Q^{\alpha }R^{\beta },  \label{eqPV2}
\end{equation}%
which is antisymmetric under interchange of $\mu $ and $\nu $, and
the full
lepton tensor for the helicity-difference matrix elements may be written%
\begin{equation}
\chi _{\mu \nu }^{\mathrm{PV,hel.\,diff.}}=a_{A}\chi _{\mu \nu
}^{(s)}+a_{V}\chi _{\mu \nu }^{(a)},  \label{eqPV3}
\end{equation}%
where the Standard Model electroweak couplings are%
\begin{eqnarray}
a_{V} &=&-(1-4\sin ^{2}\theta _{W})\simeq -0.092  \label{eqPV4} \\
a_{A} &=&-1  \label{eqPV5}
\end{eqnarray}%
using for the weak mixing angle the value $\sin ^{2}\theta
_{W}\simeq 0.227$.

The hadronic (proton) tensor is similar: the symmetric tensor is the
analog
of the unpolarized tensor given in Eq.\ (\ref{eq69})%
\begin{equation}
\widetilde{W}_{(s)}^{\mu \nu }=-\widetilde{W}_{1}\left( g^{\mu \nu }-\frac{%
Q^{\mu }Q^{\nu }}{Q^{2}}\right)
+\frac{1}{m_{p}^{2}}\widetilde{W}_{2}T^{\mu }T^{\nu }  \label{eqPV6}
\end{equation}%
which has the same form as the unpolarized proton tensor, but with
different structure functions $\widetilde{W}_{1,2}$ to be discussed
below; these are differentiated by the tildes in Eq.\ (\ref{eqPV6}).
Likewise the antisymmetric proton tensor that is the analog of that
in Eq.\ (\ref{eqPV2})
is%
\begin{equation}
\widetilde{W}_{(a)}^{\mu \nu
}=\frac{i}{m_{p}^{2}}\widetilde{W}_{3}\epsilon ^{\mu \nu \alpha
^{\prime }\beta ^{\prime }}Q_{\alpha ^{\prime }}T_{\beta ^{\prime }}
\label{eqPV7}
\end{equation}%
with an additional structure function $\widetilde{W}_{3}$.

The structure functions (functions only of $\tau $) occurring above in Eq. (%
\ref{eqPV6}) are the analogs of the familiar $W_{1,2}$ in PC elastic
e-p scattering given above (see Sect. \ref{sec:tensors}) except that
the PV analogs involve interferences between the $\gamma $ and
$Z^{0}$ diagrams and so contain products of EM form factors and
their weak neutral current
counterparts (indicated with tildes; see \cite{DDS88}):%
\begin{eqnarray}
\widetilde{W}_{1} &=&\tau G_{M}^{p}\widetilde{G}_{M}^{p}  \label{eqPV7c} \\
\widetilde{W}_{2} &=&\frac{1}{1+\tau }\left[ G_{E}^{p}\widetilde{G}%
_{E}^{p}+\tau G_{M}^{p}\widetilde{G}_{M}^{p}\right] .
\label{eqPV7d}
\end{eqnarray}%
These are all of polar vector type and go with the leptonic axial
vector term (\textit{i.e.,} proportional to $a_{A}$) to make the VA
interference. The antisymmetric case has a leptonic part that is a
polar vector, but an interference of one polar and one axial vector
contribution for the proton
to make the VA interference (see \cite{DDS88}):%
\begin{equation}
\widetilde{W}_{3}=-\frac{1}{2}G_{M}^{p}\widetilde{G}_{A}^{p}.
\label{eqPV7e}
\end{equation}%
Finally, the weak neutral current form factors in the Standard Model
are the
following%
\begin{eqnarray}
\widetilde{G}_{E}^{p} &=&\frac{1}{2}\left[ (1-4\sin ^{2}\theta
_{W})G_{E}^{p}-G_{E}^{n}-G_{E}^{(s)}\right]  \label{eqPV7f} \\
\widetilde{G}_{M}^{p} &=&\frac{1}{2}\left[ (1-4\sin ^{2}\theta
_{W})G_{M}^{p}-G_{M}^{n}-G_{M}^{(s)}\right]  \label{eqPV7g} \\
\widetilde{G}_{A}^{p} &=&\frac{1}{2}\left[
G_{A}^{(1)}-G_{A}^{(s)}\right] , \label{eqPV7h}
\end{eqnarray}%
where now the neutron's electromagnetic form factors $G_{E,M}^{n}$
enter, there are potential strangeness form factors
$G_{E,M,A}^{(s)}$ and the axial
vector, isovector form factor $G_{A}^{(1)}$ (that, for instance, enters in $%
n\rightarrow p$ $\beta $-decay) occurs in Eq.\ (\ref{eqPV7h}).

When contracting these leptonic and hadronic tensors, of course only
the symmetric or antisymmetric contributions contract with one
another and no cross terms (symmetric times antisymmetric) can
occur. The results are the
following: for the contraction of the symmetric tensors we have%
\begin{eqnarray}
\chi _{\mu \nu }^{(s)}\widetilde{W}_{(s)}^{\mu \nu } &=&-Q^{2}\widetilde{W}%
_{1}+2m_{p}^{2}\left[ \widetilde{\epsilon }^{2}-2\tau \widetilde{\epsilon }%
-\tau \right] \widetilde{W}_{2}  \label{eqPV8} \\
&=&\frac{1}{2}V_{0}\left( \widetilde{W}_{2}+2\widetilde{W}_{1}\tan
^{2}\theta _{e}^{\prime }/2\right) ,  \label{eqPV9}
\end{eqnarray}%
where $\widetilde{\epsilon }$ is defined via Eq.\ (\ref{eq74cx}) and
the last equality comes from using Eq.\ (\ref{eq74gg}); these
developments completely parallel the unpolarized contraction
discussed in Sect. \ref{sec:contract}. Note that the result here
again involves the definition of an effective angle, \textit{i.e.,}
Eq.\ (\ref{eq74hh}). For the contraction of the
antisymmetric tensors the result is%
\begin{eqnarray}
\chi _{\mu \nu }^{(a)}\widetilde{W}_{(a)}^{\mu \nu } &=&-8m_{p}^{2}\tau (%
\widetilde{\epsilon }-\tau )\widetilde{W}_{3}  \label{eqPV10} \\
&=&\frac{1}{2}V_{0}\frac{\tau (\widetilde{\epsilon }-\tau )}{\widetilde{%
\epsilon }^{2}-2\tau \widetilde{\epsilon }-\tau }\left( -2\widetilde{W}%
_{3}\right) .  \label{eqPV11}
\end{eqnarray}%
These results are for a general frame and thus may be used directly
in the collider frame.

Using the explicit results for the ERL$_{e}$ the PV asymmetry may be written%
\begin{equation}
\mathcal{A}_{PV}=\mathcal{A}_{PV}^{0}\frac{\mathcal{N}_{PV}}{\mathcal{D}_{PV}%
},  \label{eqPV12}
\end{equation}%
where%
\begin{equation}
\mathcal{A}_{PV}^{0}=\frac{G|Q^{2}|}{2\pi \alpha \sqrt{2}}=\frac{\sqrt{2}%
G\tau }{\pi \alpha }\simeq 6.33423\times 10^{-4}\,\tau
\label{eqPV13}
\end{equation}%
with $G$ the Fermi weak coupling constant. The denominator in Eq.\ (\ref%
{eqPV12}) comes from the helicity sum cross section, namely, the PC
elastic electron scattering cross section discussed above. Omitting
factors that are
common to the numerator it may be written%
\begin{equation}
\mathcal{D}_{PV}=\mathcal{E}^{\prime}\left( G_{E}^{p}\right)
^{2}+\tau \left( G_{M}^{p}\right) ^{2},  \label{eqPV14}
\end{equation}%
namely, proportional to the familiar result involving the EM form
factors (see Sect. \ref{sec:contract}). The numerator arises from the
PV helicity
difference cross section using the contractions developed above:%
\begin{eqnarray}
\mathcal{N}_{PV}&=&a_{A}\left\{
\mathcal{E}^{\prime}G_{E}^{p}\widetilde{G}_{E}^{p}+\tau
G_{M}^{p}\widetilde{G}_{M}^{p}\right\}\nonumber\\
&& + a_{V}\left\{ \sqrt{1-{\mathcal{E}^{\prime}}^{2}}%
\sqrt{\tau (1+\tau )}G_{M}^{p}\widetilde{G}_{A}^{p}\right\} . 
\label{eqPV16}
\end{eqnarray}%
Finally, the expressions above may be checked (see \cite{DDS88}) by
going to the lab.\ frame, where for the ERL$_{e}$
$\mathcal{E}^{\prime}\to\mathcal{E}$.


\section{Typical Results for e-p Scattering in Collider Kinematics\label%
{sec:numerics}}

In presenting results in this section we have made two choices for
specific kinematics, a high-energy choice that may be typical of a
future EIC facility and a somewhat lower-energy one where it may be
possible to make measurements of the proton EM form factors at a
level of precision that is interesting (see below). The two choices
are listed in Table I along with rest-frame variables corresponding
to typical collider-frame scattering angles, $\theta_e =1^o$ and
$5^o$ --- see the discussions below.

\begin{table}
\begin{tabular}{lll} \hline & {\bf Kinematics I} & {\bf
Kinematics II} \\ \hline
$k$ (GeV/c) & 10 & 2 \\
$p$ (GeV/c) & 250 & 50 \\
$k_{rest}$ (GeV/c) & 5329 & 213.2 \\
$\theta _{e}^{rest}$ (deg) at $\theta _{e}=1^{\circ}$ & 0.00188 & 0.00938 \\
$\tan (\theta _{e}^{rest}/2)$ at $\theta _{e}=1^{\circ}$ & 1.638$\times
10^{-5}$
& 8.187$\times 10^{-5}$ \\
$1-\mathcal{E}$ at $\theta _{e}=1^{\circ}$ & 5.410$\times 10^{-10}$ & 1.341$%
\times 10^{-8}$ \\
$\theta _{e}^{rest}$ (deg) at $\theta _{e}=5^{\circ}$ & 0.00939 & 0.0469 \\
$\tan (\theta _{e}^{rest}/2)$ at $\theta _{e}=5^{\circ}$ & 8.193$\times
10^{-5}$
& 4.096$\times 10^{-4}$ \\
$1-\mathcal{E}$ at $\theta _{e}=5^{\circ}$ & 1.633$\times 10^{-8}$ & 3.385$%
\times 10^{-7}$ \\ \hline%
\end{tabular}
\caption{Selected kinematics and rest-frame
variables.\label{table1}}
\end{table}

For use in assessing the feasibility of performing asymmetry
measurements one frequently employs the so-called Figure-of-Merit
(FOM):
\begin{equation}
\mathcal{F}\equiv \left[ \frac{d\sigma }{d\Omega _{e}}\right] _{ep}^{unpol,%
\text{\textit{collider}}}\left( \left[ A\right] _{ep}^{pol,\text{\textit{%
collider}}}\right) ^{2}.  \label{eq154}
\end{equation}%
Also, to incorporate the fact that the solid angle goes as $\sin
\theta _{e}$, it is also appropriate to consider the product
$\mathcal{F}\sin \theta _{e}$. Given luminosity $L$, electron
polarization $p_e$, proton polarization $p_p$, run time $T$ and an
averaged FOM times solid angle
\begin{equation}
\mathcal{F}_{avg} \Delta\Omega_e \equiv \int_{\theta_0}^{\theta_e}
\mathcal{F} \sin \theta^{\prime\prime}_e d\sin
\theta^{\prime\prime}_e , \label{eq154a}
\end{equation}
where $\theta_0$ is whatever one wishes to choose for the lower
limit of integration over scattering angle (written in the integrand
above as $\theta^{\prime\prime}_e$ to avoid confusion with
$\theta^{\prime}_e$ which has been used above and has a different
meaning) and where the full $2\pi$ integration in azimuthal angle has been
assumed, the fractional uncertainty in the $L$ or $S$ asymmetry is
given by
\begin{equation}
f = \left\{ p_e p_p \sqrt{LT\mathcal{F}_{avg} \Delta\Omega_e}
\right\}^{-1} . \label{eq154b}
\end{equation}
For the results to follow we have assumed the conditions listed in
Table II. For simplicity the lower limit chosen for the averaging of
the FOM has been fixed to $\theta_0 = 0^o$ for the results shown in
the figures; the time assumed is relatively short. In any practical
situation these two parameters will need to be adjusted to find the
optimal choices. The present study is not intended to be more than a
preliminary exploration of typical results and so no attempt has
been made to optimize the choices here. On the other hand, a general
program, Brasil2011, has been developed to study $\vec{e}$-$\vec{p}$
scattering in collider kinematics. The code, together with a
description and sample input/output are available \footnotemark[1]. All
of the parameters above can be chosen by the user. In addition, the
program provides the option of choosing between two models for
computing the nucleon form factors, namely, a simple model which
gives a reasonable starting point for e-p studies and the vector
meson dominance plus pQCD GKex model which yields good agreement
with most of the World form factor data (see \cite{Cra10} and
references cited therein for details). Alternatively, it is
straightforward for any user to provide other form factor
representations. In the following we show selected results using
this code with the GKex form factors.

\begin{table}
\begin{tabular}{ll}
$L=$ & 10$^{34}$ cm$^{-2}$ s$^{-1}$ \\
$p_e=$ & 90\% \\
$p_p=$ & 70\% \\
$T=$ & 1000 hr \\
$\theta_0=$ & $0^o$ \\
\end{tabular}
\caption{Assumed experimental conditions.\label{table2}}
\end{table}


\begin{figure}[hbt]
  \includegraphics[width=0.47\textwidth,viewport=4 6 237 148]{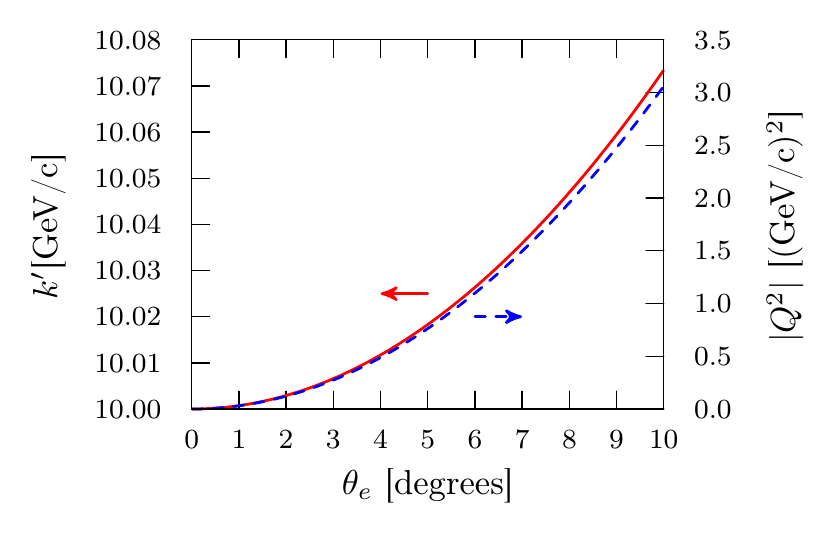}
  \caption{(color online) Kinematics I: scattered electron 3-momentum $k'$ and
  4-momentum transfer squared $|Q^2|$ versus collider-frame electron scattering angle $\theta_e$.}
  \label{fig:fig6}
\end{figure}


In Fig.\ \ref{fig:fig6} the scattered electron's 3-momentum $k'$
(left axis) and $|Q^2|$ (right axis) are shown as functions of the
electron scattering angle for kinematics I; the range of angles
chosen here is dictated by where the FOM is significant, as seen
later in Fig.\ \ref{fig:fig8}. Clearly $k'$ does not vary
significantly over the chosen range of angles, less than 75 MeV/c
out of about 10 GeV/c. For the same range of angles the scattered
proton 3-momentum goes from 250 GeV/c at $\theta_e = 0^o$ down by
only about 75 MeV/c at $10^o$ while scattering through an angular
range of $\theta_p = 0^o$ at $\theta_e = 0^o$ to $\theta_p = 0.4^o$
at $\theta_e = 10^o$. It is clear from our initial exploratory
studies that energy resolution alone is rather unlikely to be enough
to separate elastic events from inelastic ones (the code Brasil2011
contains some kinematic developments where the final hadronic state
can be taken to have invariant mass $W\neq m_p$) and some
final-state particle identification will likely be needed to isolate
elastic scattering in practical situations.


\begin{figure}[hbt]
  \includegraphics[width=0.48\textwidth,viewport=4 6 237 148]{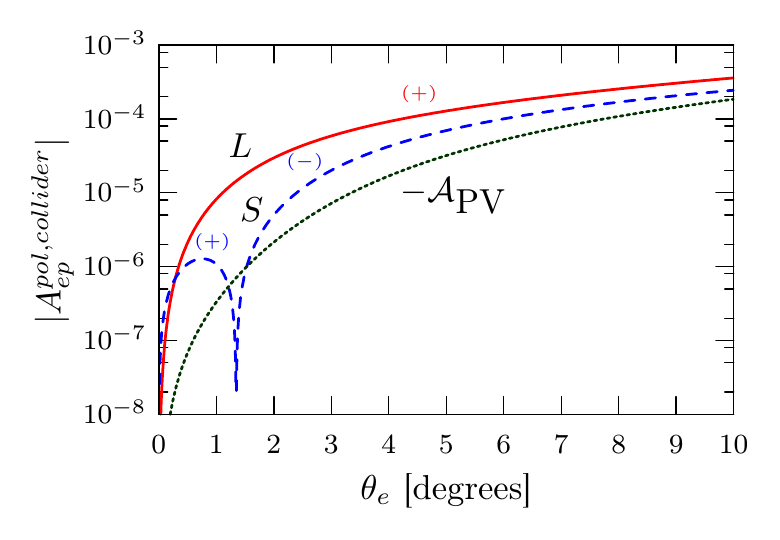}
  \caption{(color online) Kinematics I: double-polarization asymmetries
  $\left[ A\right] _{ep}^{pol,\text{\textit{collider}}}$ for $L$ and $S$
  proton polarizations versus electron scattering angle $\theta_e$.
  Also shown is the parity-violating e-p elastic asymmetry $\mathcal{A}_{PV}$
  for the same kinematics.}
  \label{fig:fig7}
\end{figure}


In Fig.\ \ref{fig:fig7} the asymmetries are shown for the same range
of electron scattering angles used above. The $\vec{e}$-$\vec{p}$
$L$ polarization is larger in magnitude than for the $S$ case,
although the two are comparable. The latter changes sign at roughly
$\theta_e = 1.3^o$. In the region where the FOM peaks (roughly
$\theta_e\sim 2^o$; see Fig.\ \ref{fig:fig8}) the $L$ asymmetry is a
few times 10$^{-5}$. One might be surprised that this is so low. The
reason is clear upon examining the rest-frame variables in Table I:
$\tan (\theta_{e}^{rest}/2)$ goes from $1.6\times 10^{-5}$ at
$\theta_e = 1^o$ to $8.2\times 10^{-5}$ at $\theta_e = 5^o$. Since
the asymmetries are proportional either to the generalized
``Rosenbluth'' factor $V_{T'}$ for $L$ polarization or to $V_{TL'}$
for $S$ polarization (see Sect.\ \ref{sec:cross-asym}) and these both
have an overall factor $\tan (\theta_{e}^{rest}/2)$ in the rest
frame \cite{Don86}, the kinematical factors are small. In other
words, the scattering in the equivalent rest frame occurs at such
small angles that the double-polarization asymmetries are
suppressed. For comparison the parity-violating asymmetry
$\mathcal{A}_{PV}$ is also shown in Fig.\ \ref{fig:fig7}. In this
case, even though the weak interaction is involved, some of the
contributions occurring in the ratio forming the asymmetry (see
Sect.\ \ref{sec:PV}) are not suppressed by similar factors. In the
same notation used above the Rosenbluth factors $V_L$ and $V_T$
occur for the VV hadronic contributions (those involving $G_{E,M}^p
{\tilde G}_{E,M}^p$ in Eq.\ (\ref{eqPV16})) and these do not contain
overall factors of $\tan (\theta_{e}^{rest}/2)$. On the other hand,
the VA interference in Eq.\ (\ref{eqPV16}) (involving $G_M^p {\tilde
G}_A^p$) has a factor $\sqrt{1-{\mathcal{E}}^2}$ and so is strongly
suppressed (see Table I).


\begin{figure}[hbt]
  \includegraphics[width=0.47\textwidth,viewport=4 6 237 148]{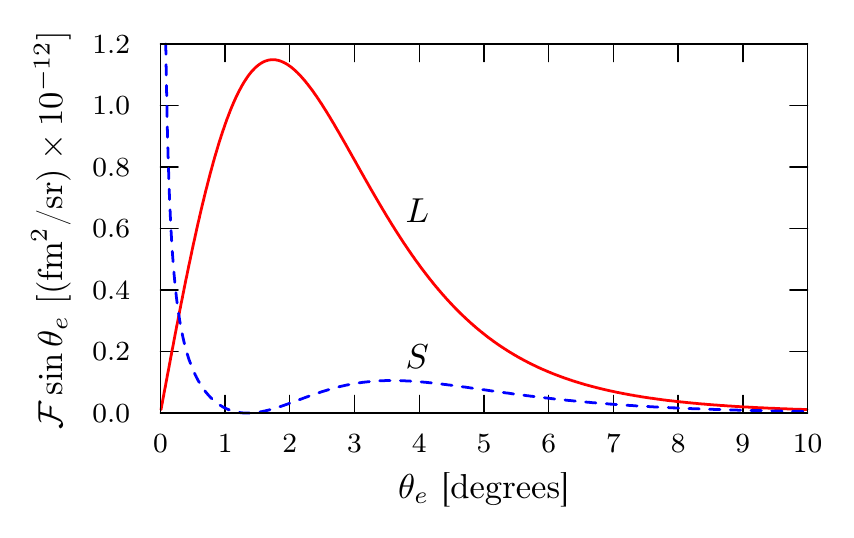}
  \caption{(color online) Kinematics I: $\mathcal{F}\sin \theta_e$
  versus electron scattering angle $\theta_e$.}
  \label{fig:fig8}
\end{figure}


Another observation from Table I is that the smallness of
$1-\mathcal{E}$, going from $5.4\times 10^{-10}$ at $\theta_e = 1^o$
to $1.6\times 10^{-8}$ at $\theta_e = 5^o$, implies something very
different about any potential $2\gamma$ corrections to the dominantly
$1\gamma$ diagram. Namely, from treatments of the former ({\it
e.g.,} see \cite{Gut11}) one expects the $2\gamma$ contributions to
vanish when $\mathcal{E} \to 1$, which is surely the case for the
rest-frame-equivalent conditions studied here.


\begin{figure}[hbt]
  \includegraphics[width=0.47\textwidth,viewport=4 6 237 148]{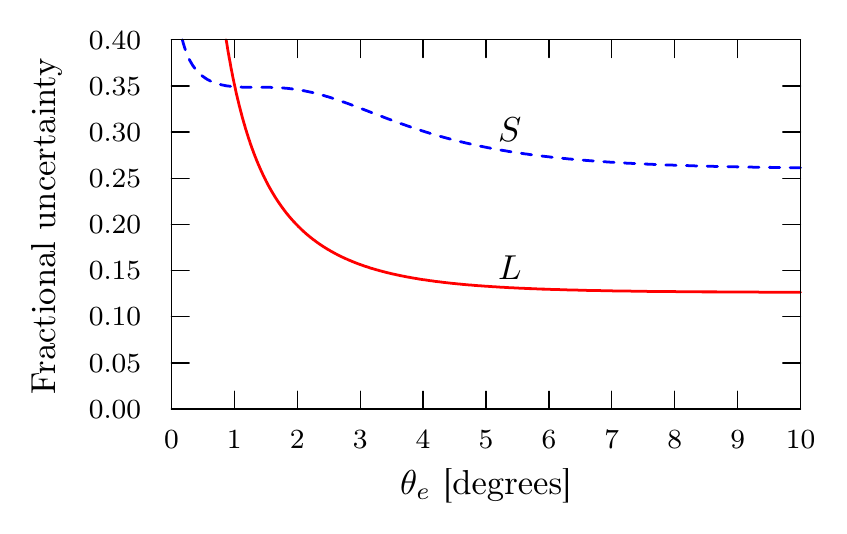}
  \caption{(color online) Kinematics I: fractional uncertainty $f$
  (see Eq.\ (\ref{eq154b}) for polarizations $L$ and $S$ versus
  electron scattering angle $\theta_e$.}
  \label{fig:fig9}
\end{figure}


Figure \ref{fig:fig9} shows the fractional uncertainty obtained
using Eq.\ (\ref{eq154b}) and implies that in a time $T=1000$ hr one
could achieve roughly a 15--20\% determination of the $L$
polarization asymmetry. Since the fractional uncertainty goes as
$1/\sqrt{T}$, clearly with longer run times even higher precision
can be obtained.

\begin{figure}[hbt]
  \includegraphics[width=0.47\textwidth,viewport=4 6 237 148]{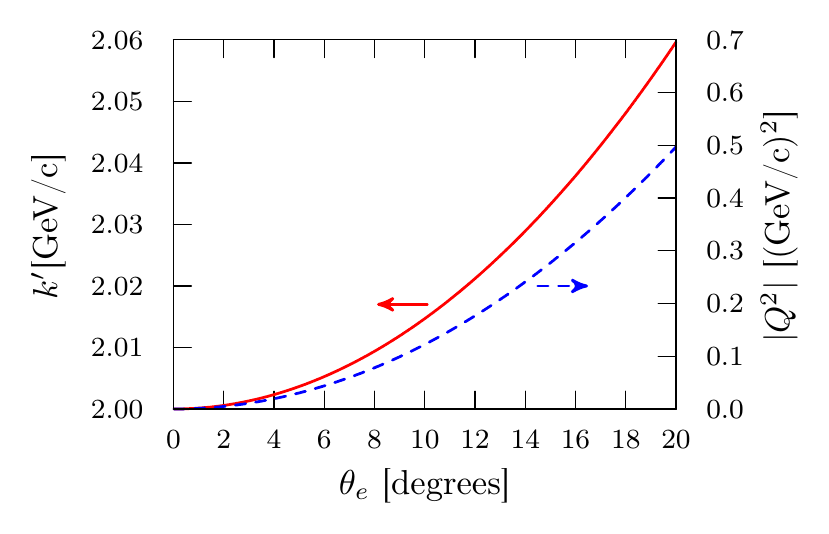}
  \caption{(color online) As for Fig.\ \ref{fig:fig6}, but now for kinematics II.}
  \label{fig:fig10}
\end{figure}

\begin{figure}[hbt]
  \includegraphics[width=0.47\textwidth,viewport=4 6 237 148]{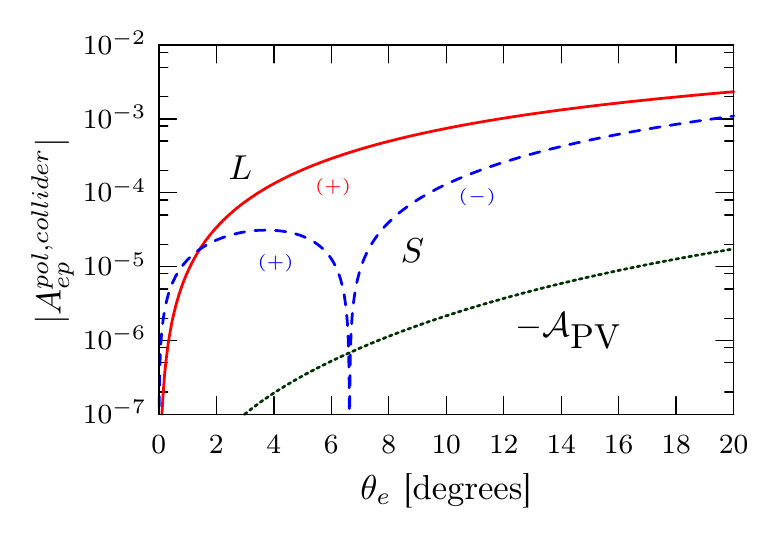}
  \caption{(color online) As for Fig.\ \ref{fig:fig7}, but now for kinematics II.}
  \label{fig:fig11}
\end{figure}

\begin{figure}[hbt]
  \includegraphics[width=0.48\textwidth,viewport=4 6 237 148]{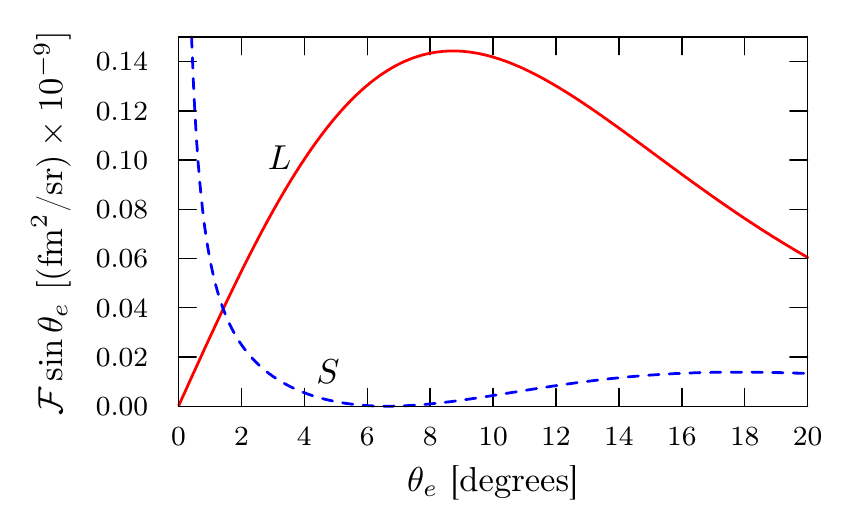}
  \caption{(color online) As for Fig.\ \ref{fig:fig8}, but now for kinematics II.}
  \label{fig:fig12}
\end{figure}

\begin{figure}[hbt]
  \includegraphics[width=0.45\textwidth,viewport=4 6 237 148]{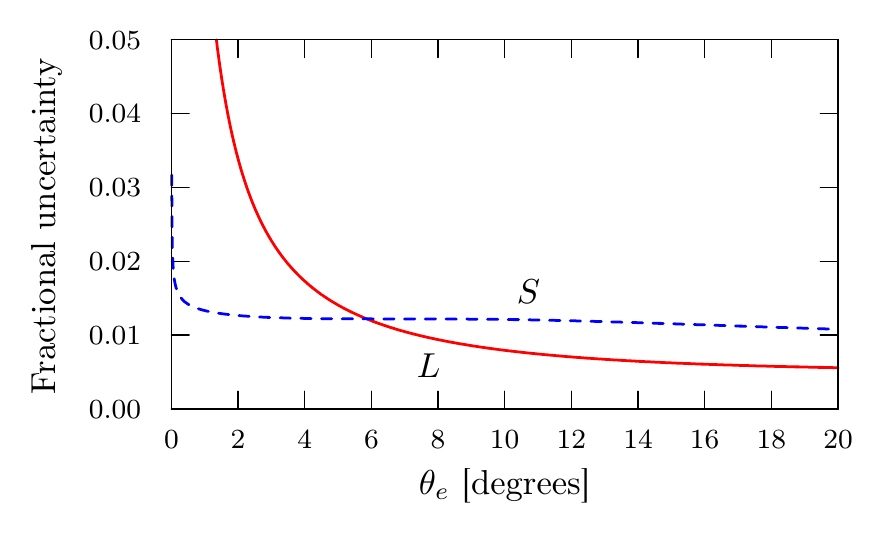}
  \caption{(color online) As for Fig.\ \ref{fig:fig9}, but now for kinematics II.}
  \label{fig:fig13}
\end{figure}

For kinematics II the results are shown in Figs. \ref{fig:fig10},
\ref{fig:fig11}, \ref{fig:fig12} and \ref{fig:fig13}. Now the
angular range is larger ($\theta_e$ up to $20^o$), again as dictated
by the significance of the FOM (see Fig.\ \ref{fig:fig12} below). The
scattered proton's 3-momentum $k'$ in this case varies from 50 GeV/c
at $\theta_e = 0^o$ to 15 MeV/c less than this at $10^o$, while the
proton's scattering angle $\theta_p$ goes from $0^o$ up to $0.4^o$
over the same range. The asymmetries are shown in Fig.
\ref{fig:fig11}: these are significantly larger than was the case for
kinematics I and now lie typically two or more orders of magnitude
above the PV asymmetry. Figure \ref{fig:fig12} shows the FOM and
indicates that for the $L$ polarization case a peak occurs for
$\theta_e$ between $8^o$ and $9^o$. This yields the fractional
uncertainty displayed in Fig.\ \ref{fig:fig13}, clearly showing that
a 1--2\% determination of the asymmetries may be possible, given the
assumed luminosity, polarizations and run time.


\section{Conclusions}
\label{sec:concl}

In this study elastic $\vec{e}$-$\vec{p}$ cross sections and
asymmetries have been considered in collider kinematics. The
formalism has been developed directly using the electron and proton
tensors and working in a general frame --- since the resulting
expressions are covariant it is easy to evaluate the results in any
chosen frame, including the system where the proton is at rest and
where the asymmetries and cross sections are well known, thereby
providing a sensitive check on the formalism. In context,
parity-violating elastic $\vec{e}$-$p$ scattering has also been
explored in collider kinematics to be able to compare and contrast
the resulting asymmetry with the double-polarization
(parity-conserving) results.

Several observations and conclusions can be made from these studies:

\begin{itemize}

\item The double-polarization asymmetries are relatively small,
since the effective rest-frame electron scattering angle is very
small for typical collider kinematics and since both the $L$ and $S$
asymmetries in the rest frame have an overall factor of $\tan
(\theta_{e}^{rest}/2)$. In contrast, the PV asymmetry has
contributions that do not contain this factor and therefore survive
when the electron scattering angle becomes very small. Indeed, for
some kinematical situations the PV asymmetry is only about one order
of magnitude smaller than the PC asymmetries.

\item The $L$ and $S$ asymmetries, while small, are still
sufficiently large that it may prove possible to make relatively
high-precision measurements of them with a future EIC facility as it
is presently envisioned. If, with such a facility, it proves possible to
use beams that have a large dynamical range, including lower
energies and beams that are more symmetrical in energy (the EIC
designs being considered are typically asymmetric with the proton
beam being much higher in energy than the electron beam), then very
likely quite high-quality determinations of the double-polarization
asymmetries can be made.

\item Given that the double-polarization asymmetries can be measured
with sufficient precision two paths may be followed: (1) the proton
EM form factors themselves may be studied; and given that we know
these form factors reasonably well from fixed-target experiments,
(2) the asymmetries may be used to determine the product of the
electron and proton polarizations when the focus is on other e-p
reactions, including studies of DIS.

\item With respect to point (1) above, the collider kinematics are
very unusual in that the effective rest-frame kinematics typically
occur at very large energies and very small angles. This means that
the virtual photon longitudinal polarization $\mathcal{E}$ is
extremely close to unity where it is predicted that $2\gamma$
corrections to the dominantly $1\gamma$ diagram go to zero.

\item Specifics of how the proton EM form factors enter the
asymmetries are interesting: if the anomalous magnetic or Pauli form
factor $F_2^p$ were zero, as is the case for a point Dirac particle,
then the $L/S$ structure of the asymmetries would be quite
different. The fact that $F_2^p \neq 0$ leads to clear signatures in
the double-polarization asymmetries.

\item Finally, while this work is a theoretical study and has been focused
on the formalism plus presentations of a few typical results, some
initial exploration has been made of the issues that will probably
confront any practical experiment. In particular, it is very
unlikely for high-energy collider kinematics that energy resolution
alone will be capable of isolating elastic e-p scattering from
inelastic scattering. Instead one will have to detect both the
scattered electron and specify the final hadronic state including
the elastic events where only the scattered proton occurs. This is
not the point of the present study, although a computer code has
been written and is available \footnotemark[1] for others involved in
design studies for a future EIC facility.
\end{itemize}

\appendix 
\section *{Appendix}
The conventions of Bjorken and Drell \cite{BjD64} are used throughout
together with the follow notation: 4-vectors are written with capital
letters
\begin{eqnarray}
A^{\mu } &=&(A^{0},A^{1},A^{2},A^{3})=(A^{0},\mathbf{a})  \label{eq1} \\
A_{\mu } &=&(A^{0},-A^{1},-A^{2},-A^{3})=(A^{0},-\mathbf{a}),  \label{eq2}
\end{eqnarray}%
where 3-vectors are written with bold lowercase letters and their magnitudes
with normal lowercase letters, $a=|\mathbf{a}|$. The scalar product of two
4-vectors $A$ and $B$ is given by
\begin{equation}
A\cdot B=A^{\mu }B_{\mu }=A^{0}B^{0}-\mathbf{a}\cdot \mathbf{b,}  \label{eq3}
\end{equation}%
where the summation convention is employed, namely repeated Greek indices
are summed $(\mu =0,1,2,3)$. Thus one has for the scalar product of a
4-vector with itself
\begin{equation}
A^{2}=\left( A^{0}\right) ^{2}-a^{2}.  \label{eq4}
\end{equation}%
Throughout we use $\hslash =c=1$.

\begin{acknowledgments}

This work was supported in part by NSF Award 0754425 through the
Hampton University UnIPhy-REU program and the MSRP program at MIT
(CS) and in part by the U.S. Department of Energy under cooperative
agreement DE-FC02-94ER40818 (TWD). The authors also wish to thank R.
Milner and J. Bernauer of MIT for helpful discussions during the
course of this study.

\end{acknowledgments}


\bibliography{collider}

\end{document}